\documentclass[journal=nalefd,manuscript=letter]{achemso}
\usepackage[version=3]{mhchem} % Formula subscripts using \ce{}
\usepackage[T1]{fontenc}       % Use modern font encodings
\usepackage{pdfpages}
\newcommand{\exciting}{{\usefont{T1}{lmtt}{b}{n}exciting}}
\author{Wahib Aggoune}
\affiliation{Institut f\"ur Physik and IRIS Adlershof, Humboldt-Universit\"at zu Berlin, 12489 Berlin, Germany}
\alsoaffiliation{Laboratoire de Physique Th\'eorique, Facult\'e des Sciences Exactes, Universit\'e de Bejaia, 06000 Bejaia, Algeria}
\author{Caterina Cocchi}
\affiliation{Institut f\"ur Physik and IRIS Adlershof, Humboldt-Universit\"at zu Berlin, 12489 Berlin, Germany}
\alsoaffiliation{European Theoretical Spectroscopic Facility (ETSF)}
\email{caterina.cocchi@physik.hu-berlin.de}
\author{Dmitrii Nabok}
\affiliation{Institut f\"ur Physik and IRIS Adlershof, Humboldt-Universit\"at zu Berlin, 12489 Berlin, Germany}
\alsoaffiliation{European Theoretical Spectroscopic Facility (ETSF)}
\author{Karim Rezouali}
\affiliation{Laboratoire de Physique Th\'eorique, Facult\'e des Sciences Exactes, Universit\'e de Bejaia, 06000 Bejaia, Algeria}
\author{Mohamed Akli Belkhir}
\affiliation{Laboratoire de Physique Th\'eorique, Facult\'e des Sciences Exactes, Universit\'e de Bejaia, 06000 Bejaia, Algeria}
\author{Claudia Draxl}
\affiliation{Institut f\"ur Physik and IRIS Adlershof, Humboldt-Universit\"at zu Berlin, 12489 Berlin, Germany}
\alsoaffiliation{European Theoretical Spectroscopic Facility (ETSF)}
\title{Enhanced Light-Matter Interaction in Graphene/h-BN van der Waals Heterostructures}
\begin{document}
%%%%%%%%%%%%%%%%%%%%%%%%%%%%%%%%%%%%%%%%%%%%%%%%%%%%%%%%%%%%%%%%%%%%%
\newpage

\begin{abstract}
By investigating the optoelectronic properties of prototypical graphene/hexagonal boron nitride (h-BN) heterostructures, we demonstrate how a nanostructured combination of these materials can lead to a dramatic enhancement of light-matter interaction and give rise to unique excitations. 
In the framework of \textit{ab initio} many-body perturbation theory, we show that such heterostructures absorb light over a broad frequency range, from the near-infrared to the ultraviolet (UV), and that each spectral region is characterized by a specific type of excitations.
Delocalized electron-hole pairs in graphene dominate the low-energy part of the spectrum, while strongly bound electron-hole pairs in h-BN are preserved in the near-UV.
Besides these features, characteristic of the pristine constituents, charge-transfer excitations appear across the visible region.
Remarkably, the spatial distribution of the electron and the hole can be selectively tuned by modulating the stacking arrangement of the individual building blocks.
Our results open up unprecedented perspectives in view of designing van der Waals heterostructures with tailored optoelectronic features.
\end{abstract}
%%%%%%%%%%%%%%%%%%%%%%%%%%%%%%%%%%%%%%%%%%%%%%%%%%%%%%%%%%%%%%%%
\section*{Graphical TOC Entry}
\begin{figure}%
\centering
\includegraphics[height=5cm]{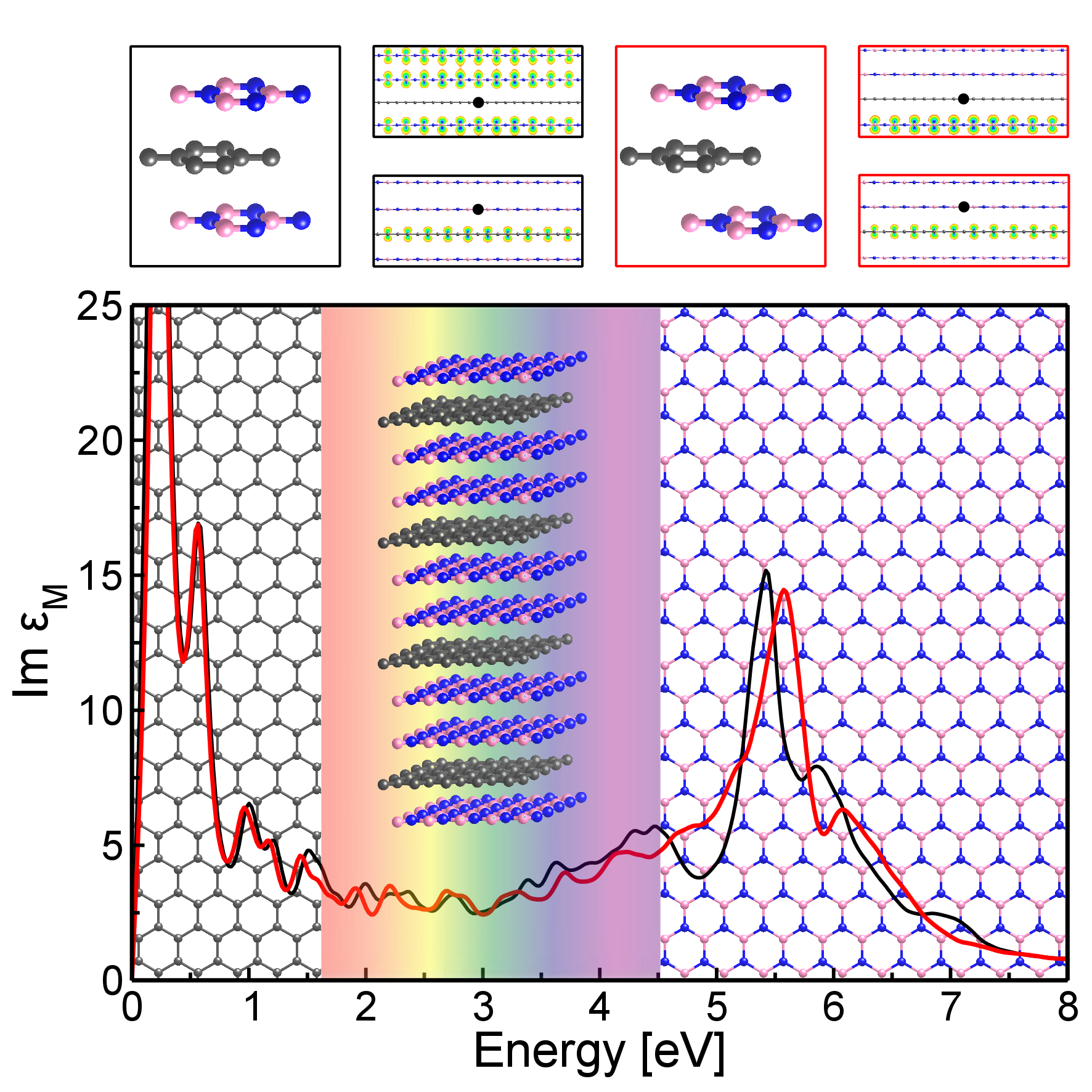}
\end{figure}

%\begin{tocentry}
%\centering
%\includegraphics[height=5cm]{../Figures_revision/Abstract_JPCL.png}
%\end{tocentry}

\textbf{Keywords}: 2D materials, excitons, many-body perturbation theory, UV-vis spectra

%%%%%%%%%%%%%%%%%%%%%%%%%%%%%%%%%%%%%%%%%%%%%%%%%%%%%%%%%%%%%%%%%%%%%
\newpage
%*******************************
%  INTRODUCTION
%*******************************
%\section{Introduction}
Van der Waals (vdW) heterostructures are a new frontier of materials science \cite{geim-grig13nat}.
The possibility of stacking atomically-thin layers with nanoscale precision has opened unprecedented opportunities to create materials with customized characteristics. 
New properties can be accessed through a combination of the constituents, which maintain their intrinsic features \cite{osad-sasa12am,xu+13cr}.
This perspective is particularly appealing in the field of optoelectronics.
The response of materials to electromagnetic radiation, consisting in a multitude of diversified phenomena, is extremely sensitive to their atomic structure and, consequently, to their electronic properties.
Systems efficiently absorbing over a broad frequency range can be designed through an engineered stacking of single layers.
Charge-transfer excitations, with the electron and the hole delocalized on different layers, can be created at the interface \cite{hong+14natn,ceba+14nano,rive+15natcom,lati+15prb,haas+16prb,lati+17nl}.
At the same time, the spatial separation of the electron-hole pairs can be enhanced through a systematic modulation of the structural properties \cite{bern+13nl,fang+14pnas,with+15natm}. 

Graphene and hexagonal boron nitride (h-BN) monolayers are ideal candidates to achieve this goal, as they exhibit a complementary behavior when interacting with light.
The optical spectrum of graphene, a peculiar semi-metal \cite{neto+09rmp}, is dominated by a zero-energy resonance, while it is rather featureless in the visible region \cite{yang+09prl}.
Collective excitations, such as excitons and plasmons, occur only at ultraviolet (UV) frequencies \cite{kram+08prl,trev+10prb,desp+13prb,wach+13prb}.
On the other hand, h-BN is a large band-gap material transparent to visible light, which exhibits unique optical properties in the near-UV range \cite{wata+04natm,kubo+07sci} related to the presence of strongly bound excitons \cite{arna+06prl,wirt+08prl,gala+11prb}.
Combining these two materials to form a vdW heterostructure enables one to benefit from their individual characteristics and to access new features \cite{saka+15jpsj}.
The small mismatch between their lattices has already triggered a number of pioneering studies in this direction \cite{brit+12sci,brit+12nl,mish+14natn,li+16small}, further boosted by the opportunity to exploit h-BN in view of opening a band gap in graphene \cite{giov+07prb,dean+10natn,mayo+11nl}. 
When interacting with light, graphene/h-BN vdW heterostructures are expected to show all their potential \cite{gao+12nl}, as recently demonstrated also for plasmonic excitations \cite{zhan+14afm,woes+15natm,jia+15acsp}.

In this work, we study the optoelectronic properties of prototypical periodic graphene/h-BN heterostructures.
Through a detailed analysis of the spectra, enabled by a highly precise state-of-the-art \textit{ab initio} many-body approach, we demonstrate that different types of excitations coexist in such a system, each of them dominating a well-defined frequency range.
We focus on charge-transfer excitations that are created at the interface and show that a selective modulation of the stacking of h-BN layers with respect to each other and to graphene can tune the spatial distribution of the electron-hole ($e$-$h$) pairs.
As such, our results pave the way for selectively enhancing light-matter interaction through nano-patterning. 

\begin{figure*}
\centering
\includegraphics[width=.95\textwidth]{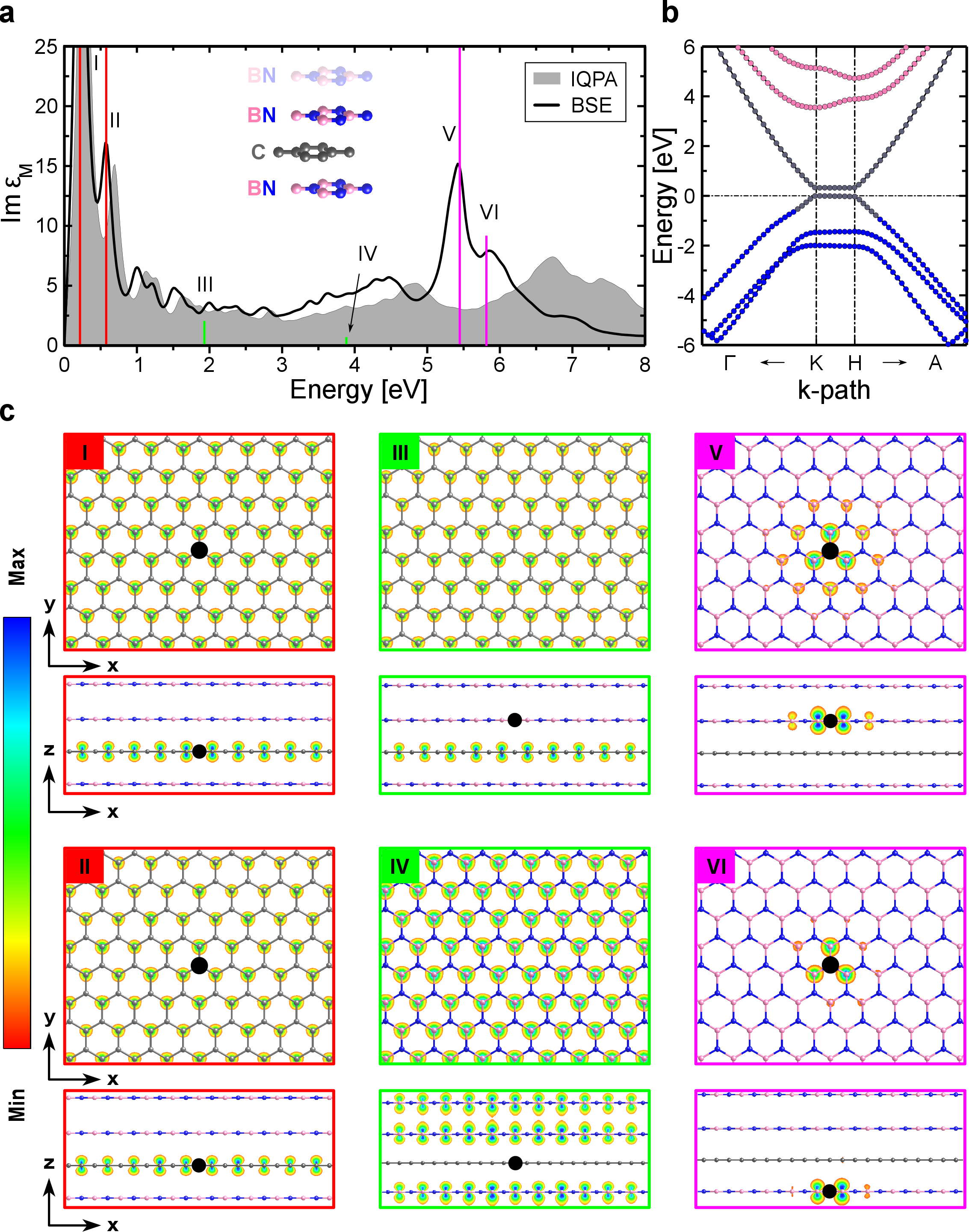}%
\caption{\textbf{a}: In-plane component of the imaginary part of the macroscopic dielectric function of a B-g-B stacked h-BN/graphene/h-BN heterostructure (inset), computed by including (BSE, solid line) and by neglecting (IQPA, shaded area) excitonic effects. \textbf{b}: Quasi-particle electronic structure of the system. The band character is indicated using the color code of the atomic species (C: gray, N: blue, B: pink). The valence-band maximum is set to zero. \textbf{c}: Two-dimensional projections of the electron component of the $e$-$h$ wave-functions highlighted in the spectrum in panel \textbf{a}. The corresponding position of the hole is marked by a black dot.}
\label{fig:B-g-B}
\end{figure*}

%
%*******************************
%  SYSTEMS
%*******************************
%
To perform this study we consider a periodic heterostructure with a graphene sheet sandwiched between four h-BN layers.
Such a system is modeled by a trilayer unit cell (h-BN/graphene/h-BN), infinitely replicated by means of periodic boundary conditions.
In this configuration, each h-BN layer directly interacts with both graphene and another h-BN sheet.
An important parameter is represented by the stacking arrangement of the layers in the in-plane directions. 
Here we focus on a structure where h-BN layers are displaced with respect to graphene such that the N atoms are located on \textit{hollow} sites with respect to the hexagonal carbon lattice (Fig. \ref{fig:B-g-B}a, inset).
We refer to this configuration as the B-g-B stacking, where g stands for graphene, and B indicates the displacement of the h-BN layers, with respect to the carbon sheet, according to the conventional nomenclature for stacked systems \cite{bern24prsla}.
More details on the structural properties of the heterostructure are provided in the Supporting Information (SI).

The electronic and optical properties of the B-g-B heterostructure are summarized in Fig. \ref{fig:B-g-B}.
The interaction with h-BN, although weak, is sufficient to open a quasi-particle gap of 250 meV in graphene along the K-H path of the Brillouin zone (BZ).
This result is in good agreement with previous studies based on density-functional theory \cite{rama+11nl,zhon+12ns,saka+14prb} and many-body perturbation theory \cite{khar-saro11nl,quhe+12npgam,hues+13prb} performed on analogous systems.
On the other hand, the interaction with graphene tends to reduce the intrinsic band gap of h-BN, which in this heterostructure amounts to 5.15 eV at the high-symmetry point K (see Fig.~\ref{fig:B-g-B}b). 
This value is about 0.5 eV smaller than the quasi-particle gap reported for bulk boron nitride in the corresponding stacking sequence \cite{bour+14acsp,note-gap}. 
Such a band-gap reduction is mainly ascribed to polarization effects, which have been extensively discussed for molecules and polymers adsorbed on graphene \cite{neat+06prl,pusc+12prb,fu+17pccp}, and which turn out to be non negligible even in the case of interacting monolayer insulators such as h-BN and carbon fluoride \cite{fu+16jpcc}.
The electronic properties of the heterostructure are reflected in the optical spectrum (Fig \ref{fig:B-g-B}a), as well as in the wealth of excitations dominating specific energy regions (Fig \ref{fig:B-g-B}c).
Among them, we immediately identify three main types, depending on their spatial extension along the in-plane and out-of-plane directions: delocalized excitations with both the electron and the hole in the same (I and II) or in different layers (III and IV), as well as localized excitons in the h-BN layer (V and VI).
In the following, we discuss in detail each spectral region and the excitation types characterizing it.

The low-frequency part of the spectrum, up to 1 eV, is dominated by interband transitions within the graphene layer, consistent with optical-conductivity measurements \cite{mak+08prl,pere10rmp}.
Due to the symmetry of the honeycomb carbon lattice, the first peak (I) and its shoulder (II) are double-degenerate (more details in the SI, Fig. S4a).
These excitations exhibit $\pi$-$\pi^*$ character, as shown in Fig. \ref{fig:B-g-B}c by the corresponding correlated probability of finding the electron for a fixed hole position. 
Specifically, the first absorption maximum (I) at 220 meV comes from a transition along the K-H path of the BZ, while II has major contributions in the vicinity of K, where the electronic wave-functions of valence-band maximum (VBM) and conduction-band minimum (CBM) are located on inequivalent carbon atoms in the unit cell (see also Figs. S2a and S5).
These intense peaks at IR frequencies are generated by the finite band-gap induced by the interaction of the carbon monolayer with the neighboring h-BN sheets \cite{giov+07prb}.  Consequently, the zero-energy divergence dominating the absorption spectrum of graphene, as an effect of its semimetallic character \cite{yang+09prl,trev+10prb}, disappears in the heterostructure, with the peak being blue-shifted due to the presence of the quasi-particle gap. 
\bibnote{In graphene the excitation energies of peaks I and II converge to the same value. Such a behavior is expected also in this system, but not perfectly reproduced due to the enhanced numerical complexity of the superlattice. The character of these excitations is nonetheless consistent with the existing literature \cite{yang+09prl,trev+10prb}, to which we refer for a detailed analysis of the optical spectrum of graphene. Thus, the essence of our results is unaffected.}

The interband character of the graphene-derived excitations is confirmed by the appearance of analogous absorption maxima in the spectrum computed within the independent quasi-particle approximation (IQPA). The $e$-$h$ interaction does not affect the spectral shape nor the nature of the excitations, characterized by a delocalized $\pi^*$-like distribution of the electron in one inequivalent carbon atom in the unit cell, with the hole being located on the other carbon atom within the graphene layer (Fig. \ref{fig:B-g-B}c). Excitonic effects essentially red-shift the peak position by 30 meV  with respect to the gap, which coincides with the onset of the IQPA spectrum at 250 meV.
Also the peculiar nature of the highest occupied band, with C-like character in the vicinity of K-H and with N-like character elsewhere in the BZ, is directly determined by the periodic alternation of the monolayers.
In the UV region, between 5 and 6 eV, we find the spectroscopic signatures of h-BN, mainly related to the excitonic peak centered at 5.4 eV.
A double-degenerate bound exciton dominates the spectrum of the bulk material \cite{arna+06prl,wirt+08prl,gala+11prb}, with its characteristics being preserved in the heterostructure.
In the latter case, the exciton binding energy cannot be quantified with respect to the fundamental gap, which is given by the graphene bands (Fig. \ref{fig:B-g-B}b).
However, comparison between the BSE and the IQPA spectra indicates that the two main excitations (V and VI) stem from the manifold of interband transitions between 6 and 8 eV (Fig. \ref{fig:B-g-B}a).
Considering the absorption maximum of the broad hump between 6 and 8 eV in the IQPA spectrum, we estimate the binding energy of excitation V to be of the order of 1 eV.
In this region, the dipole coupling between the single-particle transitions generates spatially confined $e$-$h$ pairs with large oscillator strength (Fig. \ref{fig:B-g-B}c).
At the same time, the $e$-$h$ Coulomb interaction shifts the spectral weight to lower energy.
The correlated electron distribution of these excitons is confined within a triangular region that spreads over a few unit cells around the fixed hole position, in analogy with the excitations predicted for bulk h-BN \cite{arna+06prl,wirt+08prl,gala+11prb}. 
In Fig. \ref{fig:B-g-B}c we report the corresponding averaged densities, while individual plots are shown in the SI (Fig. S4c).
The electron and the hole sit on the same layer, making both V and VI intralayer excitons.
This property comes directly from the character of the electronic states contributing to these excitations. 
Along the K-H path, the N-like VBM-1 and VBM-2, as well as the B-like states above the CBM, namely CBM+1 and CBM+2, are uniformly distributed on both h-BN layers in the unit cell (Fig. S2a).
The stacking arrangement of neighboring boron nitride sheets thus directly influences their interaction.
For this reason, regardless of the position of the hole in both excitons, the corresponding electron is always found in the same h-BN layer.
The peak centered at 5.4 eV is formed by a number of excitations with the same character as V (Fig. S3a). 
On the other hand, the weaker peak at 6 eV, in addition to exciton VI, embraces several excitations with rather mixed character, corresponding to delocalized \textit{e-h} pairs.

In the visible and near-UV range, where the spectra of the constituents are rather featureless, we find the actual fingerprints of the heterostructure.
The oscillator strength between 1.6 eV and 5 eV is weak compared to other spectral regions, owing to the charge-transfer character of most of the excitations that take place in this frequency window.
With the hole located on h-BN and the electron on the carbon layer, or vice versa, the wave-function overlap is significantly lower than in the in-plane excitations discussed above, but still non zero, thanks to the $\pi$-$\pi^*$ interaction between the layers.
In the visible region, between approximately 1.6 eV and 3.2 eV, the excitations mainly stem from transitions from the N-like VBM-1/VBM-2 to the graphene-like CBM.
An exemplary excitation of this kind is the one labeled by III (double degenerate, see Fig. S4a), which contributes to the peak at 1.9 eV, as shown in Fig. \ref{fig:B-g-B}a. 
Regardless of whether the hole is fixed on h-BN above or below graphene, the electron distribution has $\pi^*$ character, and is spread over the carbon sheet. 
At higher energies, between 3.5 and 5 eV, we find $e$-$h$ pairs with the hole stemming from the graphene VBM and the electron promoted to the boron-like CBM+1 and CBM+2.
As discussed above, the wave-functions associated with the B-like CBM+1 and CBM+2 are spread over all h-BN layers.
For this reason, the electron distribution of the charge-transfer excitation IV (also double degenerate, see Fig. S4b) is delocalized in both in-plane and out-of-plane directions (Fig. \ref{fig:B-g-B}c). 
The presence of the corresponding peaks also in the IQPA spectrum indicates the interband nature of both III and IV.
Similar to the graphene-related $\pi$-$\pi^*$ transitions (I and II), excitonic effects push these peaks to lower energies.
In this case, the red-shift amounts to about 0.1 eV. 
The relatively weak binding of these charge-transfer excitations is expected to ease the dissociation of the corresponding $e$-$h$ pairs.

\begin{figure*}
\centering
\includegraphics[width=.95\textwidth]{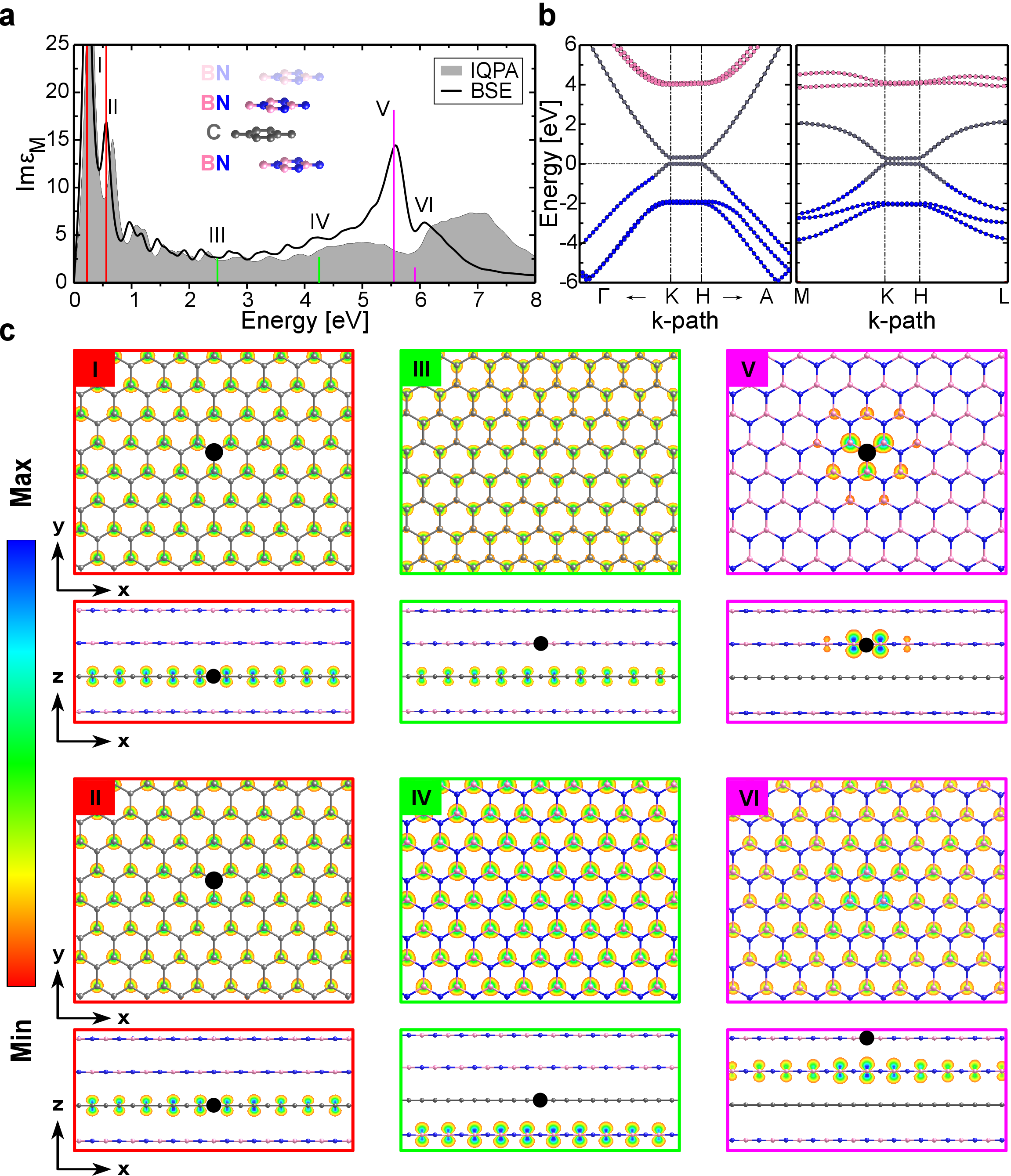}%
\caption{\textbf{a}: In-plane component of the imaginary part of the macroscopic dielectric function of a B-g-C stacked h-BN/graphene/h-BN heterostructure (inset), computed by including (BSE, solid line) and by neglecting (IQPA, shaded area) excitonic effects. \textbf{b}: Quasi-particle electronic structure of the system in the vicinity of the high-symmetry points K and H, approached from two different directions. The band character is indicated using the color code of the atomic species (C: gray, N: blue, B: pink). The valence-band maximum is set to zero. \textbf{c}:  Two-dimensional projections of the electron component of the $e$-$h$ wave-functions highlighted in the spectrum in panel \textbf{a}. The corresponding position of the hole is marked by a black dot.
}
\label{fig:B-g-C}
\end{figure*}

The spatial extension of the excitations can be further tuned by modifying the stacking arrangement of the boron nitride layers with respect to the carbon sheet.
For this purpose, we consider another heterostructure with B-g-C stacking arrangement.
In this case the two h-BN layers in the unit cell are laterally displaced with respect to each other and to graphene. 
In this way, the boron atom of one h-BN sheet and the nitrogen atom of the other one are at the hollow site of the carbon honeycomb lattice (Fig. \ref{fig:B-g-C}a, inset).
Overall, the spectrum shown in Fig. \ref{fig:B-g-C}a does not exhibit significant differences compared to its counterpart in the B-g-B heterostructure.
Clearly, also the quasi-particle band structure is similar to the one of the B-g-B system.
In Fig. \ref{fig:B-g-C}b we plot the band dispersion in the vicinity of the high-symmetry points K and H, approached from two different directions (see also Fig. S1e).
This is relevant for the later discussion about the excitation character.
A band gap of about 260 meV is opened in graphene due to its interaction with the neighboring boron nitride layers, while the energy separation between the highest occupied and lowest unoccupied bands with N- and B-like character, respectively, is approximately 6 eV. 
As in the case of the B-g-B heterostructure discussed above, the quasi-particle gap of h-BN is reduced by polarization effects due to the interaction with graphene.
The renormalization with respect to the bulk material with the same stacking sequence amounts to 0.1 eV \cite{bour+14acsp,note-gap}.
The relative displacement of the h-BN layers with respect to each other affects the distribution along the K-H path of the electronic wave-functions in both the valence and the conduction region \cite{liu+03prb}.
In fact, in the B-g-C heterostructure the VBM-1 and the VBM-2 as well as the CBM+1 and the CBM+2 are separated by about 80 meV along the K-H path. 
The corresponding electronic wave-functions are thus distributed only on one individual h-BN sheet in the unit cell, and exhibit either an N-like (valence) or a B-like (conduction) character (see also Fig. S2b in the SI).

These peculiar electronic characteristics significantly influence the interlayer excitations involving h-BN layers, namely the charge-transfer transitions in the visible/near-UV region, as well as the excitons above 5 eV.
The excitations in the near-IR, dominated by the graphene $\pi$-$\pi^*$ transitions (I and II), are obviously unaffected by the stacking arrangement.
We examine first the visible/near-UV region where the charge-transfer excitations, III and IV, appear.
To be consistent with the previous analysis of the B-g-B heterostructure, we inspect here the analogous features. 
Excitation III at about 2.5 eV is again double-degenerate (Fig. S4a) and stems from transitions between the N-like VBM-2 to the graphene-like CBM (Fig. S6 in the SI).
The electron is uniformly delocalized on graphene, exhibiting a $\pi^*$-like distribution, with the hole located on the h-BN layer directly above the carbon sheet, where the probability density of VBM-2 is the maximal (see Fig. S2b in the SI). Likewise, excitation IV is given by an $e$-$h$ pair generated by transitions from the highest-occupied graphene band, VBM, and the boron-like CBM+2.
In this case, the hole is on the carbon layer, and the electron is distributed over the h-BN sheet directly below it (Fig. \ref{fig:B-g-C}c).
A few hundreds of meV lower and higher in energy compared to III and IV other excitations are present (not shown), exhibiting again interlayer nature but different $e$-$h$ distribution, consistent with the character of the h-BN bands (Fig. S2b).
A difference with respect to the B-g-B heterostructure should be noted at this point.
In the latter system, the electron distribution of excitation III is in graphene, regardless of whether the hole sits on the h-BN layer directly above or below it.
Instead in the B-g-C stacked system the correlated hole probability associated to excitation III is limited to one boron nitride sheet with the corresponding electron being again spread over the carbon lattice.
Likewise, in this heterostructure, the electron distribution of excitation IV is restricted only to one h-BN layer, with the correlated hole sitting on graphene.
These results indicate that the distribution of the $e$-$h$ pairs appear to be very sensitive to the stacking sequence.

In a similar fashion, also the h-BN excitons are significantly affected by the layer stacking.
In this case, even the shape of the corresponding peak in the spectrum is modified compared to the B-g-B heterostructure.
The absorption maximum, centered at 5.6 eV, is formed by a number of excitations (Fig. S3b) with the same character as the double-degenerate exciton V, \textit{i.e.}, exhibiting an intralayer character, analogous to their counterparts in the B-g-B stacking (Fig. \ref{fig:B-g-C}c).
However, different from the B-g-B heterostructure, exciton V in the B-g-C stacked system arises from transitions between the N-like VBM and the B-like CBM+1 away from the K-H path in the direction of the high-symmetry point M (see Fig. S6).
In addition, a charge-transfer exciton between the h-BN sheets (VI) appears at 6 eV, within a peak formed by several excitations with mixed composition and spatial delocalization.
Due to its interlayer character, VI has lower intensity compared to the in-plane exciton V.
With the hole being located on one h-BN layer, the correlated electron distribution of exciton VI is delocalized over the other inequivalent h-BN sheet in the unit cell, with an in-plane envelope modulation embracing about 14 unit cells in real space (see Fig. \ref{fig:B-g-C}c), this exciton arises from transitions between the N-like VBM-1 and the B-like CBM+1 along the K-H path (see Fig. S6 in the SI), which
are distributed on the h-BN sheet below and above graphene, respectively (Fig. S2b). 
Such a feature appears only in the B-g-C heterostructure, due to the lateral displacement of the two h-BN layers in the unit cell.
The interplay between stacking arrangement and electronic wave-function delocalization evidently affects also ``pure'' h-BN excitations.
Although less relevant from a technological viewpoint, this result is an additional confirmation of the potential tunability of the optical properties of graphene/h-BN heterostructures.

The graphene/h-BN heterostructures considered in this work are semiconductors exhibiting a \textit{straddling} band alignment.
This \textit{type I} electronic configuration is optimal for stimulated emission, and is thus exploited for applications such as lasers and light-emitting diodes (LEDs) \cite{lau-zory93book}. 
A prototypical LED formed by vdW-stacked graphene, h-BN, and \ce{MoS2} layers was indeed produced recently through a well-defined arrangement of metallic, insulating and direct-gap seminconducting monolayers, respectively \cite{with+15natm}. 
Here, in a simpler system, consisting of a graphene/h-BN vdW heterostructure, we have demonstrated that further modulation of the electronic and optical properties can be achieved by selectively varying the stacking displacements of the h-BN sheets.

Before concluding, it is worth considering the results presented in this work in the context of the existing research on graphene/h-BN heterostructures. 
One of the main aspects to take into account is the slight lattice mismatch between the two materials, which triggers the formation of Moir{\'e} patterns \cite{xue+11natm,yank+12natp,pono+13nat,wood+14natp,yank+14jpcm,chen+14natcom,ni+15natm,gao+15natcom}. 
Depending on the growth conditions, specifically whether either graphene or h-BN acts as a substrate, different patterns can be obtained \cite{leve+16jctc}.
Locally, the stacking arrangement may vary substantially and even coincide with the sequences considered here.
The formation of Moir{\'e} patterns affects also optical properties \cite{sach+11prb,sanj+14prb,much+13prb,wall+13prb,wall+13prb1,vanw+14prl,slot+15prl,wall+15ap}.
New spectral features may appear above the absorption onset, with decreasing intensity at increasing magnitude of the mismatch angle \cite{aber+13njp}. 
The heterostructures considered in this work, with a perfect lattice matching between the individual monolayers, are obviously idealized structures. Nonetheless, the analysis of their optical fingerprints presented here is an essential step forward in view of identifying and understanding the features of real systems, where not only Moir{\'e}-like superlattices, but also buckling and distortions, typically appear and impact light-matter interaction processes.

%****************************
%  CONCLUSIONS
%****************************
To summarize, we have shown that periodic graphene/h-BN heterostructures absorb electromagnetic radiation over an extended frequency range, going from the near-IR to the UV.
While the electronic and optical properties of the constituents are essentially preserved, the interaction driven by layer stacking promotes new features in the visible/near-UV region.
Due to the interaction with the neighboring h-BN sheets, a gap is opened in graphene, with intense $\pi$-$\pi^*$ interband transition below 1 eV.
In the visible window and beyond, between 1.6 eV and 5 eV, a number of weakly bound charge-transfer excitations appear, with the corresponding electron and hole distributed on either graphene or h-BN.
The $e$-$h$ separation can be selectively tuned by modifying the stacking arrangement, that impacts the wave-function overlap, and thus leads to an increased absorption intensity.
The versatile interplay between structural, electronic, and photo-response properties in graphene/h-BN heterostructures make such materials ideal candidates and an exceptional playground for optoelectronics.

%*******************************
%  METHODS
%*******************************
\section*{Theoretical Methods and Computational Details}
Ground-state properties are computed in the framework of density functional theory (DFT), within the generalized gradient approximation for the exchange-correlation functional (Perdew-Burke-Ernzerhof parameterization \cite{PBE}).
The DFT-D2 approach proposed by S. Grimme \cite{grim06jcc} is adopted to account for van der Waals interactions between the layers.
Optical spectra are obtained in the framework of many-body perturbation theory.
Quasi-particle (QP) energies are computed within the $G_0W_0$ approximation \cite{hedi65pr,hybe-loui85prl}. 
Optical spectra are obtained from the solution of the Bethe-Salpeter equation (BSE), an effective two-body equation for the electron-hole two-particle Green's function \cite{hank-sham80prb,stri88rnc}.
The BSE Hamiltonian reads $H^{BSE} = H^{diag} + 2H^{x} + H^{dir}$, where the first term $H^{diag}$ accounts for \textit{vertical} transitions, while the other two terms incorporate electron-hole \textit{exchange} ($H^{x}$) and the screened Coulomb interaction ($H^{dir}$). 
The excitation energies $E^{\lambda}$ are the eigenvalues of the secular equation associated to the BSE Hamiltonian: $  \sum_{v'c'\mathbf{k'}} H^{BSE}_{vc\mathbf{k},v'c'\mathbf{k'}}A^{\lambda}_{v'c'\mathbf{k'}} = E^{\lambda}A^{\lambda}_{vc\mathbf{k}}$, where $v$ and $c$ indicate valence and conduction states, respectively. 
The eigenvectors $A^{\lambda}$ provide information about the intensity of the excitation, through the oscillator strength, given by the square modulus of $ \mathbf{t}_{\lambda} = \sum_{vc\mathbf{k}} A^{\lambda}_{v c \mathbf{k}} \frac{\langle v \mathbf{k} \vert \widehat{\mathbf{p}} \vert c \mathbf{k} \rangle}{\epsilon_{c \mathbf{k}}\ -\ \epsilon_{v \mathbf{k}}}$.
Moreover, they indicate the character and the composition of excitations, being the coefficients of the two-particle wave-functions $\Psi^{\lambda}(\mathbf{r}_{e},\mathbf{r}_{h}) = \sum_{v c \mathbf{k}} A^{\lambda}_{vc\mathbf{k}}\phi_{c\mathbf{k}}(\mathbf{r}_{e})\phi^{*}_{v\mathbf{k}}(\mathbf{r}_{h})$.
Absorption spectra are represented by the imaginary part of the macroscopic dielectric function $\mathrm{Im}\varepsilon_M = \dfrac{8\pi^2}{\Omega} \sum_{\lambda} |\mathbf{t}_{\lambda}|^2 \delta(\omega - E^{\lambda})$, where $\Omega$ is the unit cell volume.

All calculations are performed using {\exciting} \cite{gula+14jpcm}, an all-electron full-potential code, implementing the family of linearized augmented planewave plus local orbitals methods.
In the ground-state calculations, a basis-set cutoff R$_{MT}$G$_{max}$=7 is used.
For all atomic species involved (C, B, and N) a muffin-tin radius R$_{MT}$=1.3 bohr is adopted.
The sampling of the Brillouin zone (BZ) is performed with a 30 $\times$ 30 $\times$ 8 $\textbf{k}$-grid. 
Both lattice constants and internal coordinates are optimized until the residual forces on each atom are smaller than 0.003 eV/\AA{}. 
Calculations of QP corrections to the Kohn-Sham eigenvalues within the $G_{0}W_{0}$ approximation \cite{nabo+16prb} include 250 empty states, and a BZ sampling with a 18 $\times$ 18 $\times$ 4  shifted $\textbf{k}$-mesh is adopted.
For the solution of the BSE \cite{sagm-ambr09pccp} within the Tamm-Dancoff approximation, a cutoff  R$_{MT}$G$_{max}$=6 and a 30 $\times$ 30 $\times$ 4 shifted $\textbf{k}$-point mesh are adopted, then interpolated onto a 60 $\times$ 60 $\times$ 4 mesh, using the so-called double grid method \cite{gille+13prb,kamm+12prb}.
In the calculation of the response function and the screened Coulomb potential 100 empty bands are included. 
In the construction and diagonalization of the BSE Hamiltonian 3 occupied and 3 unoccupied bands are considered. 
Local-field effects are taken into account, by including 41 $|\mathbf{G}+\mathbf{q}|$ vectors.
For the resulting spectra, a Lorentzian broadening of 0.1 eV is applied.
Atomic structures and isosurfaces are produced with the VESTA software \cite{momm-izum11jacr}.

Input and output files of our calculations can be downloaded from the NOMAD repository at this link: \texttt{http://dx.doi.org/10.17172/NOMAD/2017.03.16-1}

%%%%%%%%%%%%%%%%%%%%%%%%%%%%%%%%%%%%%%%%%%%%%%%%%%%%%%%%%%%%%%%%%%%%%
\begin{acknowledgement}
The authors are grateful to A. Gulans, S. Lubeck, P. Pavone, Q. Fu, and S. Tillack for fruitful discussions.
This work was partly funded by the German Research Foundation (DFG), through the Collaborative Research Centers 658 and 951.
W.A.~acknowledges financial support from the Algerian Ministry of High Education and Scientific Research.
C.C.~acknowledges financial support from the Berliner Chancengleichheitsprogramm (BCP) and from IRIS Adlershof.
\end{acknowledgement}
%%%%%%%%%%%%%%%%%%%%%%%%%%%%%%%%%%%%%%%%%%%%%%%%%%%%%%%%%%%%%%%%%%%%%
\begin{suppinfo}
Structural and electronic properties of the considered heterostructures are reported, as well as additional details on the character and the composition of the optical excitations.
\end{suppinfo}

%\bibliography{bib,notes}

\begin{mcitethebibliography}{84}
\providecommand*\natexlab[1]{#1}
\providecommand*\mciteSetBstSublistMode[1]{}
\providecommand*\mciteSetBstMaxWidthForm[2]{}
\providecommand*\mciteBstWouldAddEndPuncttrue
  {\def\EndOfBibitem{\unskip.}}
\providecommand*\mciteBstWouldAddEndPunctfalse
  {\let\EndOfBibitem\relax}
\providecommand*\mciteSetBstMidEndSepPunct[3]{}
\providecommand*\mciteSetBstSublistLabelBeginEnd[3]{}
\providecommand*\EndOfBibitem{}
\mciteSetBstSublistMode{f}
\mciteSetBstMaxWidthForm{subitem}{(\alph{mcitesubitemcount})}
\mciteSetBstSublistLabelBeginEnd
  {\mcitemaxwidthsubitemform\space}
  {\relax}
  {\relax}

\bibitem[Geim and Grigorieva(2013)Geim, and Grigorieva]{geim-grig13nat}
Geim,~A.~K.; Grigorieva,~I.~V. Van der Waals heterostructures.
  \emph{Nature~(London)~} \textbf{2013}, \emph{499}, 419--425\relax
\mciteBstWouldAddEndPuncttrue
\mciteSetBstMidEndSepPunct{\mcitedefaultmidpunct}
{\mcitedefaultendpunct}{\mcitedefaultseppunct}\relax
\EndOfBibitem
\bibitem[Osada and Sasaki(2012)Osada, and Sasaki]{osad-sasa12am}
Osada,~M.; Sasaki,~T. Two-Dimensional Dielectric Nanosheets: Novel
  Nanoelectronics From Nanocrystal Building Blocks. \emph{Adv.~Mater.~}
  \textbf{2012}, \emph{24}, 210--228\relax
\mciteBstWouldAddEndPuncttrue
\mciteSetBstMidEndSepPunct{\mcitedefaultmidpunct}
{\mcitedefaultendpunct}{\mcitedefaultseppunct}\relax
\EndOfBibitem
\bibitem[Xu \latin{et~al.}(2013)Xu, Liang, Shi, and Chen]{xu+13cr}
Xu,~M.; Liang,~T.; Shi,~M.; Chen,~H. Graphene-like two-dimensional materials.
  \emph{Chem.~Rev.~} \textbf{2013}, \emph{113}, 3766--3798\relax
\mciteBstWouldAddEndPuncttrue
\mciteSetBstMidEndSepPunct{\mcitedefaultmidpunct}
{\mcitedefaultendpunct}{\mcitedefaultseppunct}\relax
\EndOfBibitem
\bibitem[Hong \latin{et~al.}(2014)Hong, Kim, Shi, Zhang, Jin, Sun, Tongay, Wu,
  Zhang, and Wang]{hong+14natn}
Hong,~X.; Kim,~J.; Shi,~S.-F.; Zhang,~Y.; Jin,~C.; Sun,~Y.; Tongay,~S.; Wu,~J.;
  Zhang,~Y.; Wang,~F. Ultrafast charge transfer in atomically thin MoS2/WS2
  heterostructures. \emph{Nature Nanotech.} \textbf{2014}, \emph{9},
  682--686\relax
\mciteBstWouldAddEndPuncttrue
\mciteSetBstMidEndSepPunct{\mcitedefaultmidpunct}
{\mcitedefaultendpunct}{\mcitedefaultseppunct}\relax
\EndOfBibitem
\bibitem[Ceballos \latin{et~al.}(2014)Ceballos, Bellus, Chiu, and
  Zhao]{ceba+14nano}
Ceballos,~F.; Bellus,~M.~Z.; Chiu,~H.-Y.; Zhao,~H. Ultrafast charge separation
  and indirect exciton formation in a MoS2--MoSe2 van der Waals
  heterostructure. \emph{ACS~Nano} \textbf{2014}, \emph{8}, 12717--12724\relax
\mciteBstWouldAddEndPuncttrue
\mciteSetBstMidEndSepPunct{\mcitedefaultmidpunct}
{\mcitedefaultendpunct}{\mcitedefaultseppunct}\relax
\EndOfBibitem
\bibitem[Rivera \latin{et~al.}(2015)Rivera, Schaibley, Jones, Ross, Wu,
  Aivazian, Klement, Seyler, Clark, Ghimire, Yan, Mandrus, Yao, and
  Xu]{rive+15natcom}
Rivera,~P.; Schaibley,~J.~R.; Jones,~A.~M.; Ross,~J.~S.; Wu,~S.; Aivazian,~G.;
  Klement,~P.; Seyler,~K.; Clark,~G.; Ghimire,~N.~J. \latin{et~al.}
  Observation of long-lived interlayer excitons in monolayer MoSe2--WSe2
  heterostructures. \emph{Nature Comm.} \textbf{2015}, \emph{6}\relax
\mciteBstWouldAddEndPuncttrue
\mciteSetBstMidEndSepPunct{\mcitedefaultmidpunct}
{\mcitedefaultendpunct}{\mcitedefaultseppunct}\relax
\EndOfBibitem
\bibitem[Latini \latin{et~al.}(2015)Latini, Olsen, and Thygesen]{lati+15prb}
Latini,~S.; Olsen,~T.; Thygesen,~K.~S. Excitons in van der Waals
  heterostructures: The important role of dielectric screening.
  \emph{Phys.~Rev.~B} \textbf{2015}, \emph{92}, 245123\relax
\mciteBstWouldAddEndPuncttrue
\mciteSetBstMidEndSepPunct{\mcitedefaultmidpunct}
{\mcitedefaultendpunct}{\mcitedefaultseppunct}\relax
\EndOfBibitem
\bibitem[Haastrup \latin{et~al.}(2016)Haastrup, Latini, Bolotin, and
  Thygesen]{haas+16prb}
Haastrup,~S.; Latini,~S.; Bolotin,~K.; Thygesen,~K.~S. Stark shift and
  electric-field-induced dissociation of excitons in monolayer
  MoS$_2$ and hBN/MoS$_2$
  heterostructures. \emph{Phys.~Rev.~B} \textbf{2016}, \emph{94}, 041401\relax
\mciteBstWouldAddEndPuncttrue
\mciteSetBstMidEndSepPunct{\mcitedefaultmidpunct}
{\mcitedefaultendpunct}{\mcitedefaultseppunct}\relax
\EndOfBibitem
\bibitem[Latini \latin{et~al.}(2017)Latini, Winther, Olsen, and
  Thygesen]{lati+17nl}
Latini,~S.; Winther,~K.~T.; Olsen,~T.; Thygesen,~K.~S. Interlayer excitons and
  Band Alignment in MoS2/hBN/WSe2 van der Waals Heterostructures.
  \emph{Nano~Lett.~} \textbf{2017}, \emph{17}, 938--945\relax
\mciteBstWouldAddEndPuncttrue
\mciteSetBstMidEndSepPunct{\mcitedefaultmidpunct}
{\mcitedefaultendpunct}{\mcitedefaultseppunct}\relax
\EndOfBibitem
\bibitem[Bernardi \latin{et~al.}(2013)Bernardi, Palummo, and
  Grossman]{bern+13nl}
Bernardi,~M.; Palummo,~M.; Grossman,~J.~C. Extraordinary sunlight absorption
  and one nanometer thick photovoltaics using two-dimensional monolayer
  materials. \emph{Nano~Lett.~} \textbf{2013}, \emph{13}, 3664--3670\relax
\mciteBstWouldAddEndPuncttrue
\mciteSetBstMidEndSepPunct{\mcitedefaultmidpunct}
{\mcitedefaultendpunct}{\mcitedefaultseppunct}\relax
\EndOfBibitem
\bibitem[Fang \latin{et~al.}(2014)Fang, Battaglia, Carraro, Nemsak, Ozdol,
  Kang, Bechtel, Desai, Kronast, Unal, Conti, Conlon, Palsson, Marting, Minor,
  Fadley, Yablonovitch, Maboudian, and Javey]{fang+14pnas}
Fang,~H.; Battaglia,~C.; Carraro,~C.; Nemsak,~S.; Ozdol,~B.; Kang,~J.~S.;
  Bechtel,~H.~A.; Desai,~S.~B.; Kronast,~F.; Unal,~A.~A. \latin{et~al.}  Strong
  interlayer coupling in van der Waals heterostructures built from single-layer
  chalcogenides. \emph{Proc. Natl. Acad. Sci. USA} \textbf{2014}, \emph{111},
  6198--6202\relax
\mciteBstWouldAddEndPuncttrue
\mciteSetBstMidEndSepPunct{\mcitedefaultmidpunct}
{\mcitedefaultendpunct}{\mcitedefaultseppunct}\relax
\EndOfBibitem
\bibitem[Withers \latin{et~al.}(2015)Withers, Del Pozo-Zamudio, Mishchenko,
  Rooney, Gholinia, Watanabe, Taniguchi, Haigh, Geim, Tartakovskii, and
  Novoselov]{with+15natm}
Withers,~F.; Del Pozo-Zamudio,~O.; Mishchenko,~A.; Rooney,~A.; Gholinia,~A.;
  Watanabe,~K.; Taniguchi,~T.; Haigh,~S.; Geim,~A.; Tartakovskii,~A.
  \latin{et~al.}  Light-emitting diodes by band-structure engineering in van
  der Waals heterostructures. \emph{Nature Mat.} \textbf{2015}, \emph{14},
  301--306\relax
\mciteBstWouldAddEndPuncttrue
\mciteSetBstMidEndSepPunct{\mcitedefaultmidpunct}
{\mcitedefaultendpunct}{\mcitedefaultseppunct}\relax
\EndOfBibitem
\bibitem[Neto \latin{et~al.}(2009)Neto, Guinea, Peres, Novoselov, and
  Geim]{neto+09rmp}
Neto,~A.~C.; Guinea,~F.; Peres,~N.~M.; Novoselov,~K.~S.; Geim,~A.~K. The
  electronic properties of graphene. \emph{Rev.~Mod.~Phys.~} \textbf{2009},
  \emph{81}, 109\relax
\mciteBstWouldAddEndPuncttrue
\mciteSetBstMidEndSepPunct{\mcitedefaultmidpunct}
{\mcitedefaultendpunct}{\mcitedefaultseppunct}\relax
\EndOfBibitem
\bibitem[Yang \latin{et~al.}(2009)Yang, Deslippe, Park, Cohen, and
  Louie]{yang+09prl}
Yang,~L.; Deslippe,~J.; Park,~C.-H.; Cohen,~M.~L.; Louie,~S.~G. Excitonic
  effects on the optical response of graphene and bilayer graphene.
  \emph{Phys.~Rev.~Lett.~} \textbf{2009}, \emph{103}, 186802\relax
\mciteBstWouldAddEndPuncttrue
\mciteSetBstMidEndSepPunct{\mcitedefaultmidpunct}
{\mcitedefaultendpunct}{\mcitedefaultseppunct}\relax
\EndOfBibitem
\bibitem[Kramberger \latin{et~al.}(2008)Kramberger, Hambach, Giorgetti,
  R{\"u}mmeli, Knupfer, Fink, B{\"u}chner, Reining, Einarsson, Maruyama,
  Sottile, Hannewald, Olevano, Marinopoulos, and Pichler]{kram+08prl}
Kramberger,~C.; Hambach,~R.; Giorgetti,~C.; R{\"u}mmeli,~M.; Knupfer,~M.;
  Fink,~J.; B{\"u}chner,~B.; Reining,~L.; Einarsson,~E.; Maruyama,~S.
  \latin{et~al.}  Linear plasmon dispersion in single-wall carbon nanotubes and
  the collective excitation spectrum of graphene. \emph{Phys.~Rev.~Lett.~}
  \textbf{2008}, \emph{100}, 196803\relax
\mciteBstWouldAddEndPuncttrue
\mciteSetBstMidEndSepPunct{\mcitedefaultmidpunct}
{\mcitedefaultendpunct}{\mcitedefaultseppunct}\relax
\EndOfBibitem
\bibitem[Trevisanutto \latin{et~al.}(2010)Trevisanutto, Holzmann, C{\^o}t{\'e},
  and Olevano]{trev+10prb}
Trevisanutto,~P.~E.; Holzmann,~M.; C{\^o}t{\'e},~M.; Olevano,~V. Ab initio
  high-energy excitonic effects in graphite and graphene. \emph{Phys.~Rev.~B}
  \textbf{2010}, \emph{81}, 121405\relax
\mciteBstWouldAddEndPuncttrue
\mciteSetBstMidEndSepPunct{\mcitedefaultmidpunct}
{\mcitedefaultendpunct}{\mcitedefaultseppunct}\relax
\EndOfBibitem
\bibitem[Despoja \latin{et~al.}(2013)Despoja, Novko, Dekani{\'c},
  {\v{S}}unji{\'c}, and Maru{\v{s}}i{\'c}]{desp+13prb}
Despoja,~V.; Novko,~D.; Dekani{\'c},~K.; {\v{S}}unji{\'c},~M.;
  Maru{\v{s}}i{\'c},~L. Two-dimensional and $\pi$ plasmon spectra in pristine
  and doped graphene. \emph{Phys.~Rev.~B} \textbf{2013}, \emph{87},
  075447\relax
\mciteBstWouldAddEndPuncttrue
\mciteSetBstMidEndSepPunct{\mcitedefaultmidpunct}
{\mcitedefaultendpunct}{\mcitedefaultseppunct}\relax
\EndOfBibitem
\bibitem[Wachsmuth \latin{et~al.}(2013)Wachsmuth, Hambach, Kinyanjui, Guzzo,
  Benner, and Kaiser]{wach+13prb}
Wachsmuth,~P.; Hambach,~R.; Kinyanjui,~M.; Guzzo,~M.; Benner,~G.; Kaiser,~U.
  High-energy collective electronic excitations in free-standing single-layer
  graphene. \emph{Phys.~Rev.~B} \textbf{2013}, \emph{88}, 075433\relax
\mciteBstWouldAddEndPuncttrue
\mciteSetBstMidEndSepPunct{\mcitedefaultmidpunct}
{\mcitedefaultendpunct}{\mcitedefaultseppunct}\relax
\EndOfBibitem
\bibitem[Watanabe \latin{et~al.}(2004)Watanabe, Taniguchi, and
  Kanda]{wata+04natm}
Watanabe,~K.; Taniguchi,~T.; Kanda,~H. Direct-bandgap properties and evidence
  for ultraviolet lasing of hexagonal boron nitride single crystal.
  \emph{Nature Mat.} \textbf{2004}, \emph{3}, 404--409\relax
\mciteBstWouldAddEndPuncttrue
\mciteSetBstMidEndSepPunct{\mcitedefaultmidpunct}
{\mcitedefaultendpunct}{\mcitedefaultseppunct}\relax
\EndOfBibitem
\bibitem[Kubota \latin{et~al.}(2007)Kubota, Watanabe, Tsuda, and
  Taniguchi]{kubo+07sci}
Kubota,~Y.; Watanabe,~K.; Tsuda,~O.; Taniguchi,~T. Deep ultraviolet
  light-emitting hexagonal boron nitride synthesized at atmospheric pressure.
  \emph{Science} \textbf{2007}, \emph{317}, 932--934\relax
\mciteBstWouldAddEndPuncttrue
\mciteSetBstMidEndSepPunct{\mcitedefaultmidpunct}
{\mcitedefaultendpunct}{\mcitedefaultseppunct}\relax
\EndOfBibitem
\bibitem[Arnaud \latin{et~al.}(2006)Arnaud, Lebegue, Rabiller, and
  Alouani]{arna+06prl}
Arnaud,~B.; Lebegue,~S.; Rabiller,~P.; Alouani,~M. Huge excitonic effects in
  layered hexagonal boron nitride. \emph{Phys.~Rev.~Lett.~} \textbf{2006},
  \emph{96}, 026402\relax
\mciteBstWouldAddEndPuncttrue
\mciteSetBstMidEndSepPunct{\mcitedefaultmidpunct}
{\mcitedefaultendpunct}{\mcitedefaultseppunct}\relax
\EndOfBibitem
\bibitem[Wirtz \latin{et~al.}(2008)Wirtz, Marini, Gr{\"u}ning, Attaccalite,
  Kresse, and Rubio]{wirt+08prl}
Wirtz,~L.; Marini,~A.; Gr{\"u}ning,~M.; Attaccalite,~C.; Kresse,~G.; Rubio,~A.
  Comment on ``Huge excitonic effects in layered hexagonal boron nitride''.
  \emph{Phys.~Rev.~Lett.~} \textbf{2008}, \emph{100}, 189701\relax
\mciteBstWouldAddEndPuncttrue
\mciteSetBstMidEndSepPunct{\mcitedefaultmidpunct}
{\mcitedefaultendpunct}{\mcitedefaultseppunct}\relax
\EndOfBibitem
\bibitem[Galambosi \latin{et~al.}(2011)Galambosi, Wirtz, Soininen, Serrano,
  Marini, Watanabe, Taniguchi, Huotari, Rubio, and
  H{\"a}m{\"a}l{\"a}inen]{gala+11prb}
Galambosi,~S.; Wirtz,~L.; Soininen,~J.; Serrano,~J.; Marini,~A.; Watanabe,~K.;
  Taniguchi,~T.; Huotari,~S.; Rubio,~A.; H{\"a}m{\"a}l{\"a}inen,~K. Anisotropic
  excitonic effects in the energy loss function of hexagonal boron nitride.
  \emph{Phys.~Rev.~B} \textbf{2011}, \emph{83}, 081413\relax
\mciteBstWouldAddEndPuncttrue
\mciteSetBstMidEndSepPunct{\mcitedefaultmidpunct}
{\mcitedefaultendpunct}{\mcitedefaultseppunct}\relax
\EndOfBibitem
\bibitem[Sakai \latin{et~al.}(2015)Sakai, Saito, and Cohen]{saka+15jpsj}
Sakai,~Y.; Saito,~S.; Cohen,~M.~L. First-Principles Study on Graphene/Hexagonal
  Boron Nitride Heterostructures. \emph{J.~Phys.~Soc.~Jpn.~} \textbf{2015},
  \emph{84}, 121002\relax
\mciteBstWouldAddEndPuncttrue
\mciteSetBstMidEndSepPunct{\mcitedefaultmidpunct}
{\mcitedefaultendpunct}{\mcitedefaultseppunct}\relax
\EndOfBibitem
\bibitem[Britnell \latin{et~al.}(2012)Britnell, Gorbachev, Jalil, Belle,
  Schedin, Mishchenko, Georgiou, Katsnelson, Eaves, Morozov, Peres, Leist,
  Geim, Novoselov, and Ponomarenko]{brit+12sci}
Britnell,~L.; Gorbachev,~R.; Jalil,~R.; Belle,~B.; Schedin,~F.; Mishchenko,~A.;
  Georgiou,~T.; Katsnelson,~M.; Eaves,~L.; Morozov,~S. \latin{et~al.}
  Field-effect tunneling transistor based on vertical graphene
  heterostructures. \emph{Science} \textbf{2012}, \emph{335}, 947--950\relax
\mciteBstWouldAddEndPuncttrue
\mciteSetBstMidEndSepPunct{\mcitedefaultmidpunct}
{\mcitedefaultendpunct}{\mcitedefaultseppunct}\relax
\EndOfBibitem
\bibitem[Britnell \latin{et~al.}(2012)Britnell, Gorbachev, Jalil, Belle,
  Schedin, Katsnelson, Eaves, Morozov, Mayorov, Peres, Castro~Neto, Leist,
  Geim, Ponomarenko, and Novoselov]{brit+12nl}
Britnell,~L.; Gorbachev,~R.~V.; Jalil,~R.; Belle,~B.~D.; Schedin,~F.;
  Katsnelson,~M.~I.; Eaves,~L.; Morozov,~S.~V.; Mayorov,~A.~S.; Peres,~N.~M.
  \latin{et~al.}  Electron tunneling through ultrathin boron nitride
  crystalline barriers. \emph{Nano~Lett.~} \textbf{2012}, \emph{12},
  1707--1710\relax
\mciteBstWouldAddEndPuncttrue
\mciteSetBstMidEndSepPunct{\mcitedefaultmidpunct}
{\mcitedefaultendpunct}{\mcitedefaultseppunct}\relax
\EndOfBibitem
\bibitem[Mishchenko \latin{et~al.}(2014)Mishchenko, Tu, Cao, Gorbachev,
  Wallbank, Greenaway, Morozov, Morozov, Zhu, Wong, Withers, Woods, Kim,
  Watanabe, Taniguchi, Vdovin, Makarovsky, Fromhold, Fal'ko, Geim, Eaves, and
  Novoselov]{mish+14natn}
Mishchenko,~A.; Tu,~J.; Cao,~Y.; Gorbachev,~R.; Wallbank,~J.; Greenaway,~M.;
  Morozov,~V.; Morozov,~S.; Zhu,~M.; Wong,~S. \latin{et~al.}  Twist-controlled
  resonant tunnelling in graphene/boron nitride/graphene heterostructures.
  \emph{Nature Nanotech.} \textbf{2014}, \emph{9}, 808--813\relax
\mciteBstWouldAddEndPuncttrue
\mciteSetBstMidEndSepPunct{\mcitedefaultmidpunct}
{\mcitedefaultendpunct}{\mcitedefaultseppunct}\relax
\EndOfBibitem
\bibitem[Li \latin{et~al.}(2016)Li, Liu, Zhang, and Liu]{li+16small}
Li,~Q.; Liu,~M.; Zhang,~Y.; Liu,~Z. Hexagonal Boron Nitride--Graphene
  Heterostructures: Synthesis and Interfacial Properties. \emph{Small}
  \textbf{2016}, \emph{12}, 32--50\relax
\mciteBstWouldAddEndPuncttrue
\mciteSetBstMidEndSepPunct{\mcitedefaultmidpunct}
{\mcitedefaultendpunct}{\mcitedefaultseppunct}\relax
\EndOfBibitem
\bibitem[Giovannetti \latin{et~al.}(2007)Giovannetti, Khomyakov, Brocks, Kelly,
  and van~den Brink]{giov+07prb}
Giovannetti,~G.; Khomyakov,~P.~A.; Brocks,~G.; Kelly,~P.~J.; van~den Brink,~J.
  Substrate-induced band gap in graphene on hexagonal boron nitride: Ab initio
  density functional calculations. \emph{Phys.~Rev.~B} \textbf{2007},
  \emph{76}, 073103\relax
\mciteBstWouldAddEndPuncttrue
\mciteSetBstMidEndSepPunct{\mcitedefaultmidpunct}
{\mcitedefaultendpunct}{\mcitedefaultseppunct}\relax
\EndOfBibitem
\bibitem[Dean \latin{et~al.}(2010)Dean, Young, Meric, Lee, Wang, Sorgenfrei,
  Watanabe, Taniguchi, Kim, Shepard, and Hone]{dean+10natn}
Dean,~C.~R.; Young,~A.~F.; Meric,~I.; Lee,~C.; Wang,~L.; Sorgenfrei,~S.;
  Watanabe,~K.; Taniguchi,~T.; Kim,~P.; Shepard,~K. \latin{et~al.}  Boron
  nitride substrates for high-quality graphene electronics. \emph{Nature
  Nanotech.} \textbf{2010}, \emph{5}, 722--726\relax
\mciteBstWouldAddEndPuncttrue
\mciteSetBstMidEndSepPunct{\mcitedefaultmidpunct}
{\mcitedefaultendpunct}{\mcitedefaultseppunct}\relax
\EndOfBibitem
\bibitem[Mayorov \latin{et~al.}(2011)Mayorov, Gorbachev, Morozov, Britnell,
  Jalil, Ponomarenko, Blake, Novoselov, Watanabe, Taniguchi, and
  Geim]{mayo+11nl}
Mayorov,~A.~S.; Gorbachev,~R.~V.; Morozov,~S.~V.; Britnell,~L.; Jalil,~R.;
  Ponomarenko,~L.~A.; Blake,~P.; Novoselov,~K.~S.; Watanabe,~K.; Taniguchi,~T.
  \latin{et~al.}  Micrometer-scale ballistic transport in encapsulated graphene
  at room temperature. \emph{Nano~Lett.~} \textbf{2011}, \emph{11},
  2396--2399\relax
\mciteBstWouldAddEndPuncttrue
\mciteSetBstMidEndSepPunct{\mcitedefaultmidpunct}
{\mcitedefaultendpunct}{\mcitedefaultseppunct}\relax
\EndOfBibitem
\bibitem[Gao \latin{et~al.}(2012)Gao, Gao, Cannuccia, Taha-Tijerina, Balicas,
  Mathkar, Narayanan, Liu, Gupta, Peng, Yin, Rubio, and Ajayan]{gao+12nl}
Gao,~G.; Gao,~W.; Cannuccia,~E.; Taha-Tijerina,~J.; Balicas,~L.; Mathkar,~A.;
  Narayanan,~T.; Liu,~Z.; Gupta,~B.~K.; Peng,~J. \latin{et~al.}  Artificially
  stacked atomic layers: toward new van der Waals solids. \emph{Nano~Lett.~}
  \textbf{2012}, \emph{12}, 3518--3525\relax
\mciteBstWouldAddEndPuncttrue
\mciteSetBstMidEndSepPunct{\mcitedefaultmidpunct}
{\mcitedefaultendpunct}{\mcitedefaultseppunct}\relax
\EndOfBibitem
\bibitem[Zhang \latin{et~al.}(2014)Zhang, Yap, Li, Ng, Li, and Loh]{zhan+14afm}
Zhang,~K.; Yap,~F.~L.; Li,~K.; Ng,~C.~T.; Li,~L.~J.; Loh,~K.~P. Large scale
  graphene/hexagonal boron nitride heterostructure for tunable plasmonics.
  \emph{Adv.~Funct.~Mater.~} \textbf{2014}, \emph{24}, 731--738\relax
\mciteBstWouldAddEndPuncttrue
\mciteSetBstMidEndSepPunct{\mcitedefaultmidpunct}
{\mcitedefaultendpunct}{\mcitedefaultseppunct}\relax
\EndOfBibitem
\bibitem[Woessner \latin{et~al.}(2015)Woessner, Lundeberg, Gao, Principi,
  Alonso-Gonz{\'a}lez, Carrega, Watanabe, Taniguchi, Vignale, Polini, Hone,
  Hillenbrand, and Koppens]{woes+15natm}
Woessner,~A.; Lundeberg,~M.~B.; Gao,~Y.; Principi,~A.; Alonso-Gonz{\'a}lez,~P.;
  Carrega,~M.; Watanabe,~K.; Taniguchi,~T.; Vignale,~G.; Polini,~M.
  \latin{et~al.}  Highly confined low-loss plasmons in graphene--boron nitride
  heterostructures. \emph{Nature Mat.} \textbf{2015}, \emph{14}, 421--425\relax
\mciteBstWouldAddEndPuncttrue
\mciteSetBstMidEndSepPunct{\mcitedefaultmidpunct}
{\mcitedefaultendpunct}{\mcitedefaultseppunct}\relax
\EndOfBibitem
\bibitem[Jia \latin{et~al.}(2015)Jia, Zhao, Guo, Wang, Wang, and
  Xia]{jia+15acsp}
Jia,~Y.; Zhao,~H.; Guo,~Q.; Wang,~X.; Wang,~H.; Xia,~F. Tunable plasmon--phonon
  polaritons in layered graphene--hexagonal boron nitride heterostructures.
  \emph{ACS~Photon.} \textbf{2015}, \emph{2}, 907--912\relax
\mciteBstWouldAddEndPuncttrue
\mciteSetBstMidEndSepPunct{\mcitedefaultmidpunct}
{\mcitedefaultendpunct}{\mcitedefaultseppunct}\relax
\EndOfBibitem
\bibitem[Bernal(1924)]{bern24prsla}
Bernal,~J. The structure of graphite. \emph{Proc.~R.~Soc.~London~A}
  \textbf{1924}, \emph{106}, 749--773\relax
\mciteBstWouldAddEndPuncttrue
\mciteSetBstMidEndSepPunct{\mcitedefaultmidpunct}
{\mcitedefaultendpunct}{\mcitedefaultseppunct}\relax
\EndOfBibitem
\bibitem[Ramasubramaniam \latin{et~al.}(2011)Ramasubramaniam, Naveh, and
  Towe]{rama+11nl}
Ramasubramaniam,~A.; Naveh,~D.; Towe,~E. Tunable band gaps in bilayer graphene-
  BN heterostructures. \emph{Nano~Lett.~} \textbf{2011}, \emph{11},
  1070--1075\relax
\mciteBstWouldAddEndPuncttrue
\mciteSetBstMidEndSepPunct{\mcitedefaultmidpunct}
{\mcitedefaultendpunct}{\mcitedefaultseppunct}\relax
\EndOfBibitem
\bibitem[Zhong \latin{et~al.}(2012)Zhong, Amorim, Scheicher, Pandey, and
  Karna]{zhon+12ns}
Zhong,~X.; Amorim,~R.~G.; Scheicher,~R.~H.; Pandey,~R.; Karna,~S.~P. Electronic
  structure and quantum transport properties of trilayers formed from graphene
  and boron nitride. \emph{Nanoscale} \textbf{2012}, \emph{4}, 5490--5498\relax
\mciteBstWouldAddEndPuncttrue
\mciteSetBstMidEndSepPunct{\mcitedefaultmidpunct}
{\mcitedefaultendpunct}{\mcitedefaultseppunct}\relax
\EndOfBibitem
\bibitem[Sakai \latin{et~al.}(2014)Sakai, Saito, and Cohen]{saka+14prb}
Sakai,~Y.; Saito,~S.; Cohen,~M.~L. Lattice matching and electronic structure of
  finite-layer graphene/h-BN thin films. \emph{Phys.~Rev.~B} \textbf{2014},
  \emph{89}, 115424\relax
\mciteBstWouldAddEndPuncttrue
\mciteSetBstMidEndSepPunct{\mcitedefaultmidpunct}
{\mcitedefaultendpunct}{\mcitedefaultseppunct}\relax
\EndOfBibitem
\bibitem[Kharche and Nayak(2011)Kharche, and Nayak]{khar-saro11nl}
Kharche,~N.; Nayak,~S.~K. Quasiparticle band gap engineering of graphene and
  graphone on hexagonal boron nitride substrate. \emph{Nano~Lett.~}
  \textbf{2011}, \emph{11}, 5274--5278\relax
\mciteBstWouldAddEndPuncttrue
\mciteSetBstMidEndSepPunct{\mcitedefaultmidpunct}
{\mcitedefaultendpunct}{\mcitedefaultseppunct}\relax
\EndOfBibitem
\bibitem[Quhe \latin{et~al.}(2012)Quhe, Zheng, Luo, Liu, Qin, Zhou, Yu, Nagase,
  Mei, Gao, and Lu]{quhe+12npgam}
Quhe,~R.; Zheng,~J.; Luo,~G.; Liu,~Q.; Qin,~R.; Zhou,~J.; Yu,~D.; Nagase,~S.;
  Mei,~W.-N.; Gao,~Z. \latin{et~al.}  Tunable and sizable band gap of
  single-layer graphene sandwiched between hexagonal boron nitride. \emph{NPG
  Asia Materials} \textbf{2012}, \emph{4}, e6\relax
\mciteBstWouldAddEndPuncttrue
\mciteSetBstMidEndSepPunct{\mcitedefaultmidpunct}
{\mcitedefaultendpunct}{\mcitedefaultseppunct}\relax
\EndOfBibitem
\bibitem[H\"user \latin{et~al.}(2013)H\"user, Olsen, and Thygesen]{hues+13prb}
H\"user,~F.; Olsen,~T.; Thygesen,~K.~S. Quasiparticle GW calculations for
  solids, molecules, and two-dimensional materials. \emph{Phys.~Rev.~B}
  \textbf{2013}, \emph{87}, 235132\relax
\mciteBstWouldAddEndPuncttrue
\mciteSetBstMidEndSepPunct{\mcitedefaultmidpunct}
{\mcitedefaultendpunct}{\mcitedefaultseppunct}\relax
\EndOfBibitem
\bibitem[Bourrellier \latin{et~al.}(2014)Bourrellier, Amato, Galvao~Tizei,
  Giorgetti, Gloter, Heggie, March, St{\'e}phan, Reining, Kociak, and
  Zobelli]{bour+14acsp}
Bourrellier,~R.; Amato,~M.; Galvao~Tizei,~L.~H.; Giorgetti,~C.; Gloter,~A.;
  Heggie,~M.~I.; March,~K.; St{\'e}phan,~O.; Reining,~L.; Kociak,~M.
  \latin{et~al.}  Nanometric resolved luminescence in h-BN flakes: excitons and
  stacking order. \emph{ACS~Photon.} \textbf{2014}, \emph{1}, 857--862\relax
\mciteBstWouldAddEndPuncttrue
\mciteSetBstMidEndSepPunct{\mcitedefaultmidpunct}
{\mcitedefaultendpunct}{\mcitedefaultseppunct}\relax
\EndOfBibitem
\bibitem[not()]{note-gap}
A strictly quantitative comparison with the results of Ref.
  \citenum{bour+14acsp} can hardly be made, due to the different
  $\mathbf{k}$-meshes adopted for the calculations of the quasi-particle
  band-structure. In this work we use a shifted 18$\times$18$\times$4 grid,
  while in Ref. \citenum{bour+14acsp} a 6$\times$6$\times$2 mesh is adopted.
  Nonetheless, the band-gap renormalization of h-BN in presence of graphene is
  qualitatively confirmed.\relax
\mciteBstWouldAddEndPunctfalse
\mciteSetBstMidEndSepPunct{\mcitedefaultmidpunct}
{}{\mcitedefaultseppunct}\relax
\EndOfBibitem
\bibitem[Neaton \latin{et~al.}(2006)Neaton, Hybertsen, and Louie]{neat+06prl}
Neaton,~J.~B.; Hybertsen,~M.~S.; Louie,~S.~G. Renormalization of Molecular
  Electronic Levels at Metal-Molecule Interfaces. \emph{Phys.~Rev.~Lett.~}
  \textbf{2006}, \emph{97}, 216405\relax
\mciteBstWouldAddEndPuncttrue
\mciteSetBstMidEndSepPunct{\mcitedefaultmidpunct}
{\mcitedefaultendpunct}{\mcitedefaultseppunct}\relax
\EndOfBibitem
\bibitem[Puschnig \latin{et~al.}(2012)Puschnig, Amiri, and Draxl]{pusc+12prb}
Puschnig,~P.; Amiri,~P.; Draxl,~C. Band renormalization of a polymer
  physisorbed on graphene investigated by many-body perturbation theory.
  \emph{Phys.~Rev.~B} \textbf{2012}, \emph{86}, 085107\relax
\mciteBstWouldAddEndPuncttrue
\mciteSetBstMidEndSepPunct{\mcitedefaultmidpunct}
{\mcitedefaultendpunct}{\mcitedefaultseppunct}\relax
\EndOfBibitem
\bibitem[Fu \latin{et~al.}(2017)Fu, Cocchi, Nabok, Gulans, and
  Draxl]{fu+17pccp}
Fu,~Q.; Cocchi,~C.; Nabok,~D.; Gulans,~A.; Draxl,~C. Graphene-modulated
  photo-absorption in adsorbed azobenzene monolayers.
  \emph{Phys.~Chem.~Chem.~Phys.~} \textbf{2017}, \emph{19}, 6196--6205\relax
\mciteBstWouldAddEndPuncttrue
\mciteSetBstMidEndSepPunct{\mcitedefaultmidpunct}
{\mcitedefaultendpunct}{\mcitedefaultseppunct}\relax
\EndOfBibitem
\bibitem[Fu \latin{et~al.}(2016)Fu, Nabok, and Draxl]{fu+16jpcc}
Fu,~Q.; Nabok,~D.; Draxl,~C. Energy-Level Alignment at the Interface of
  Graphene Fluoride and Boron Nitride Monolayers: An Investigation by Many-Body
  Perturbation Theory. \emph{J.~Phys.~Chem.~C} \textbf{2016}, \emph{120},
  11671--11678\relax
\mciteBstWouldAddEndPuncttrue
\mciteSetBstMidEndSepPunct{\mcitedefaultmidpunct}
{\mcitedefaultendpunct}{\mcitedefaultseppunct}\relax
\EndOfBibitem
\bibitem[Mak \latin{et~al.}(2008)Mak, Sfeir, Wu, Lui, Misewich, and
  Heinz]{mak+08prl}
Mak,~K.~F.; Sfeir,~M.~Y.; Wu,~Y.; Lui,~C.~H.; Misewich,~J.~A.; Heinz,~T.~F.
  Measurement of the optical conductivity of graphene. \emph{Phys.~Rev.~Lett.~}
  \textbf{2008}, \emph{101}, 196405\relax
\mciteBstWouldAddEndPuncttrue
\mciteSetBstMidEndSepPunct{\mcitedefaultmidpunct}
{\mcitedefaultendpunct}{\mcitedefaultseppunct}\relax
\EndOfBibitem
\bibitem[Peres(2010)]{pere10rmp}
Peres,~N. Colloquium: The transport properties of graphene: An introduction.
  \emph{Rev.~Mod.~Phys.~} \textbf{2010}, \emph{82}, 2673\relax
\mciteBstWouldAddEndPuncttrue
\mciteSetBstMidEndSepPunct{\mcitedefaultmidpunct}
{\mcitedefaultendpunct}{\mcitedefaultseppunct}\relax
\EndOfBibitem
\bibitem[Not()]{Note-1}
In graphene the excitation energies of peaks I and II converge to the same
  value. Such a behavior is expected also in this system, but not perfectly
  reproduced due to the enhanced numerical complexity of the superlattice. The
  character of these excitations is nonetheless consistent with the existing
  literature \cite{yang+09prl,trev+10prb}, to which we refer for a detailed
  analysis of the optical spectrum of graphene. Thus, the essence of our
  results is unaffected.\relax
\mciteBstWouldAddEndPunctfalse
\mciteSetBstMidEndSepPunct{\mcitedefaultmidpunct}
{}{\mcitedefaultseppunct}\relax
\EndOfBibitem
\bibitem[Liu \latin{et~al.}(2003)Liu, Feng, and Shen]{liu+03prb}
Liu,~L.; Feng,~Y.; Shen,~Z. Structural and electronic properties of h-BN.
  \emph{Phys.~Rev.~B} \textbf{2003}, \emph{68}, 104102\relax
\mciteBstWouldAddEndPuncttrue
\mciteSetBstMidEndSepPunct{\mcitedefaultmidpunct}
{\mcitedefaultendpunct}{\mcitedefaultseppunct}\relax
\EndOfBibitem
\bibitem[Lau and Zory(1993)Lau, and Zory]{lau-zory93book}
Lau,~K.; Zory,~P. \emph{Ultralow threshold quantum well lasers}; San Diego, CA:
  Academic, 1993; pp 198--203\relax
\mciteBstWouldAddEndPuncttrue
\mciteSetBstMidEndSepPunct{\mcitedefaultmidpunct}
{\mcitedefaultendpunct}{\mcitedefaultseppunct}\relax
\EndOfBibitem
\bibitem[Xue \latin{et~al.}(2011)Xue, Sanchez-Yamagishi, Bulmash, Jacquod,
  Deshpande, Watanabe, Taniguchi, Jarillo-Herrero, and LeRoy]{xue+11natm}
Xue,~J.; Sanchez-Yamagishi,~J.; Bulmash,~D.; Jacquod,~P.; Deshpande,~A.;
  Watanabe,~K.; Taniguchi,~T.; Jarillo-Herrero,~P.; LeRoy,~B.~J. Scanning
  tunnelling microscopy and spectroscopy of ultra-flat graphene on hexagonal
  boron nitride. \emph{Nature Mat.} \textbf{2011}, \emph{10}, 282--285\relax
\mciteBstWouldAddEndPuncttrue
\mciteSetBstMidEndSepPunct{\mcitedefaultmidpunct}
{\mcitedefaultendpunct}{\mcitedefaultseppunct}\relax
\EndOfBibitem
\bibitem[Yankowitz \latin{et~al.}(2012)Yankowitz, Xue, Cormode,
  Sanchez-Yamagishi, Watanabe, Taniguchi, Jarillo-Herrero, Jacquod, and
  LeRoy]{yank+12natp}
Yankowitz,~M.; Xue,~J.; Cormode,~D.; Sanchez-Yamagishi,~J.~D.; Watanabe,~K.;
  Taniguchi,~T.; Jarillo-Herrero,~P.; Jacquod,~P.; LeRoy,~B.~J. Emergence of
  superlattice Dirac points in graphene on hexagonal boron nitride.
  \emph{Nature Phys.} \textbf{2012}, \emph{8}, 382--386\relax
\mciteBstWouldAddEndPuncttrue
\mciteSetBstMidEndSepPunct{\mcitedefaultmidpunct}
{\mcitedefaultendpunct}{\mcitedefaultseppunct}\relax
\EndOfBibitem
\bibitem[Ponomarenko \latin{et~al.}(2013)Ponomarenko, Gorbachev, Yu, Elias,
  Jalil, Patel, Mishchenko, Mayorov, Woods, Wallbank, Mucha-Kruczy{\'n}ski,
  Piot, Potemski, Grigorieva, Novoselov, Guinea, Fal’ko, and
  Geim]{pono+13nat}
Ponomarenko,~L.; Gorbachev,~R.; Yu,~G.; Elias,~D.; Jalil,~R.; Patel,~A.;
  Mishchenko,~A.; Mayorov,~A.; Woods,~C.; Wallbank,~J. \latin{et~al.}  Cloning
  of Dirac fermions in graphene superlattices. \emph{Nature~(London)~}
  \textbf{2013}, \emph{497}, 594--597\relax
\mciteBstWouldAddEndPuncttrue
\mciteSetBstMidEndSepPunct{\mcitedefaultmidpunct}
{\mcitedefaultendpunct}{\mcitedefaultseppunct}\relax
\EndOfBibitem
\bibitem[Woods \latin{et~al.}(2014)Woods, Britnell, Eckmann, Ma, Lu, Guo, Lin,
  Yu, Cao, Gorbachev, Kretinin, Park, Ponomarenko, Katsnelson, Gornostyrev,
  Watanabe, Taniguchi, Casiraghi, Gao, Geim, and Novoselov]{wood+14natp}
Woods,~C.; Britnell,~L.; Eckmann,~A.; Ma,~R.; Lu,~J.; Guo,~H.; Lin,~X.; Yu,~G.;
  Cao,~Y.; Gorbachev,~R. \latin{et~al.}  Commensurate-incommensurate transition
  in graphene on hexagonal boron nitride. \emph{Nature Phys.} \textbf{2014},
  \emph{10}, 451--456\relax
\mciteBstWouldAddEndPuncttrue
\mciteSetBstMidEndSepPunct{\mcitedefaultmidpunct}
{\mcitedefaultendpunct}{\mcitedefaultseppunct}\relax
\EndOfBibitem
\bibitem[Yankowitz \latin{et~al.}(2014)Yankowitz, Xue, and LeRoy]{yank+14jpcm}
Yankowitz,~M.; Xue,~J.; LeRoy,~B.~J. Graphene on hexagonal boron nitride.
  \emph{J.~Phys.~Condens.~Matter.~} \textbf{2014}, \emph{26}, 303201\relax
\mciteBstWouldAddEndPuncttrue
\mciteSetBstMidEndSepPunct{\mcitedefaultmidpunct}
{\mcitedefaultendpunct}{\mcitedefaultseppunct}\relax
\EndOfBibitem
\bibitem[Chen \latin{et~al.}(2014)Chen, Shi, Yang, Lu, Lai, Yan, Wang, Zhang,
  and Li]{chen+14natcom}
Chen,~Z.-G.; Shi,~Z.; Yang,~W.; Lu,~X.; Lai,~Y.; Yan,~H.; Wang,~F.; Zhang,~G.;
  Li,~Z. Observation of an intrinsic bandgap and Landau level renormalization
  in graphene/boron-nitride heterostructures. \emph{Nature Comm.}
  \textbf{2014}, \emph{5}, 4461\relax
\mciteBstWouldAddEndPuncttrue
\mciteSetBstMidEndSepPunct{\mcitedefaultmidpunct}
{\mcitedefaultendpunct}{\mcitedefaultseppunct}\relax
\EndOfBibitem
\bibitem[Ni \latin{et~al.}(2015)Ni, Wang, Wu, Fei, Goldflam, Keilmann,
  {\"O}zyilmaz, Neto, Xie, Fogler, and Basov]{ni+15natm}
Ni,~G.; Wang,~H.; Wu,~J.; Fei,~Z.; Goldflam,~M.; Keilmann,~F.;
  {\"O}zyilmaz,~B.; Neto,~A.~C.; Xie,~X.; Fogler,~M. \latin{et~al.}  Plasmons
  in graphene moire superlattices. \emph{Nature Mat.} \textbf{2015}, \emph{14},
  1217--1222\relax
\mciteBstWouldAddEndPuncttrue
\mciteSetBstMidEndSepPunct{\mcitedefaultmidpunct}
{\mcitedefaultendpunct}{\mcitedefaultseppunct}\relax
\EndOfBibitem
\bibitem[Gao \latin{et~al.}(2015)Gao, Song, Du, Nie, Chen, Ji, Sun, Yang,
  Zhang, and Liu]{gao+15natcom}
Gao,~T.; Song,~X.; Du,~H.; Nie,~Y.; Chen,~Y.; Ji,~Q.; Sun,~J.; Yang,~Y.;
  Zhang,~Y.; Liu,~Z. Temperature-triggered chemical switching growth of
  in-plane and vertically stacked graphene-boron nitride heterostructures.
  \emph{Nature Comm.} \textbf{2015}, \emph{6}, 6835\relax
\mciteBstWouldAddEndPuncttrue
\mciteSetBstMidEndSepPunct{\mcitedefaultmidpunct}
{\mcitedefaultendpunct}{\mcitedefaultseppunct}\relax
\EndOfBibitem
\bibitem[Leven \latin{et~al.}(2016)Leven, Maaravi, Azuri, Kronik, and
  Hod]{leve+16jctc}
Leven,~I.; Maaravi,~T.; Azuri,~I.; Kronik,~L.; Hod,~O. Inter-Layer Potential
  for Graphene/h-BN Heterostructures. \emph{J.~Chem.~Theory.~Comput.~}
  \textbf{2016}, \emph{12}, 2896--2905\relax
\mciteBstWouldAddEndPuncttrue
\mciteSetBstMidEndSepPunct{\mcitedefaultmidpunct}
{\mcitedefaultendpunct}{\mcitedefaultseppunct}\relax
\EndOfBibitem
\bibitem[Sachs \latin{et~al.}(2011)Sachs, Wehling, Katsnelson, and
  Lichtenstein]{sach+11prb}
Sachs,~B.; Wehling,~T.; Katsnelson,~M.; Lichtenstein,~A. Adhesion and
  electronic structure of graphene on hexagonal boron nitride substrates.
  \emph{Phys.~Rev.~B} \textbf{2011}, \emph{84}, 195414\relax
\mciteBstWouldAddEndPuncttrue
\mciteSetBstMidEndSepPunct{\mcitedefaultmidpunct}
{\mcitedefaultendpunct}{\mcitedefaultseppunct}\relax
\EndOfBibitem
\bibitem[San-Jose \latin{et~al.}(2014)San-Jose, Guti{\'e}rrez-Rubio, Sturla,
  and Guinea]{sanj+14prb}
San-Jose,~P.; Guti{\'e}rrez-Rubio,~A.; Sturla,~M.; Guinea,~F. Electronic
  structure of spontaneously strained graphene on hexagonal boron nitride.
  \emph{Phys.~Rev.~B} \textbf{2014}, \emph{90}, 115152\relax
\mciteBstWouldAddEndPuncttrue
\mciteSetBstMidEndSepPunct{\mcitedefaultmidpunct}
{\mcitedefaultendpunct}{\mcitedefaultseppunct}\relax
\EndOfBibitem
\bibitem[Mucha-Kruczy{\'n}ski \latin{et~al.}(2013)Mucha-Kruczy{\'n}ski,
  Wallbank, and Fal'Ko]{much+13prb}
Mucha-Kruczy{\'n}ski,~M.; Wallbank,~J.; Fal'Ko,~V. Heterostructures of bilayer
  graphene and h-BN: Interplay between misalignment, interlayer asymmetry, and
  trigonal warping. \emph{Phys.~Rev.~B} \textbf{2013}, \emph{88}, 205418\relax
\mciteBstWouldAddEndPuncttrue
\mciteSetBstMidEndSepPunct{\mcitedefaultmidpunct}
{\mcitedefaultendpunct}{\mcitedefaultseppunct}\relax
\EndOfBibitem
\bibitem[Wallbank \latin{et~al.}(2013)Wallbank, Patel, Mucha-Kruczy{\'n}ski,
  Geim, and Fal'ko]{wall+13prb}
Wallbank,~J.; Patel,~A.; Mucha-Kruczy{\'n}ski,~M.; Geim,~A.; Fal'ko,~V. Generic
  miniband structure of graphene on a hexagonal substrate. \emph{Phys.~Rev.~B}
  \textbf{2013}, \emph{87}, 245408\relax
\mciteBstWouldAddEndPuncttrue
\mciteSetBstMidEndSepPunct{\mcitedefaultmidpunct}
{\mcitedefaultendpunct}{\mcitedefaultseppunct}\relax
\EndOfBibitem
\bibitem[Wallbank \latin{et~al.}(2013)Wallbank, Mucha-Kruczy{\'n}ski, and
  Fal'ko]{wall+13prb1}
Wallbank,~J.; Mucha-Kruczy{\'n}ski,~M.; Fal'ko,~V. Moir{\'e} minibands in
  graphene heterostructures with almost commensurate 3$\times$ 3 hexagonal
  crystals. \emph{Phys.~Rev.~B} \textbf{2013}, \emph{88}, 155415\relax
\mciteBstWouldAddEndPuncttrue
\mciteSetBstMidEndSepPunct{\mcitedefaultmidpunct}
{\mcitedefaultendpunct}{\mcitedefaultseppunct}\relax
\EndOfBibitem
\bibitem[Van~Wijk \latin{et~al.}(2014)Van~Wijk, Schuring, Katsnelson, and
  Fasolino]{vanw+14prl}
Van~Wijk,~M.; Schuring,~A.; Katsnelson,~M.; Fasolino,~A. Moire patterns as a
  probe of interplanar interactions for graphene on h-BN.
  \emph{Phys.~Rev.~Lett.~} \textbf{2014}, \emph{113}, 135504\relax
\mciteBstWouldAddEndPuncttrue
\mciteSetBstMidEndSepPunct{\mcitedefaultmidpunct}
{\mcitedefaultendpunct}{\mcitedefaultseppunct}\relax
\EndOfBibitem
\bibitem[Slotman \latin{et~al.}(2015)Slotman, van Wijk, Zhao, Fasolino,
  Katsnelson, and Yuan]{slot+15prl}
Slotman,~G.; van Wijk,~M.; Zhao,~P.-L.; Fasolino,~A.; Katsnelson,~M.; Yuan,~S.
  Effect of structural relaxation on the electronic structure of graphene on
  hexagonal boron nitride. \emph{Phys.~Rev.~Lett.~} \textbf{2015}, \emph{115},
  186801\relax
\mciteBstWouldAddEndPuncttrue
\mciteSetBstMidEndSepPunct{\mcitedefaultmidpunct}
{\mcitedefaultendpunct}{\mcitedefaultseppunct}\relax
\EndOfBibitem
\bibitem[Wallbank \latin{et~al.}(2015)Wallbank, Mucha-Kruczy{\'n}ski, Chen, and
  Fal'ko]{wall+15ap}
Wallbank,~J.~R.; Mucha-Kruczy{\'n}ski,~M.; Chen,~X.; Fal'ko,~V.~I. Moir{\'e}
  superlattice effects in graphene/boron-nitride van der Waals
  heterostructures. \emph{Ann.~Phys.~} \textbf{2015}, \emph{527},
  359--376\relax
\mciteBstWouldAddEndPuncttrue
\mciteSetBstMidEndSepPunct{\mcitedefaultmidpunct}
{\mcitedefaultendpunct}{\mcitedefaultseppunct}\relax
\EndOfBibitem
\bibitem[Abergel \latin{et~al.}(2013)Abergel, Wallbank, Chen,
  Mucha-Kruczy{\'n}ski, and Fal'ko]{aber+13njp}
Abergel,~D.~S.; Wallbank,~J.; Chen,~X.; Mucha-Kruczy{\'n}ski,~M.; Fal'ko,~V.~I.
  Infrared absorption by graphene--hBN heterostructures. \emph{New.~J.~Phys.~}
  \textbf{2013}, \emph{15}, 123009\relax
\mciteBstWouldAddEndPuncttrue
\mciteSetBstMidEndSepPunct{\mcitedefaultmidpunct}
{\mcitedefaultendpunct}{\mcitedefaultseppunct}\relax
\EndOfBibitem
\bibitem[Perdew \latin{et~al.}(1996)Perdew, Burke, and Ernzerhof]{PBE}
Perdew,~J.~P.; Burke,~K.; Ernzerhof,~M. Generalized Gradient Approximation Made
  Simple. \emph{Phys.~Rev.~Lett.~} \textbf{1996}, \emph{77}, 3865--3868\relax
\mciteBstWouldAddEndPuncttrue
\mciteSetBstMidEndSepPunct{\mcitedefaultmidpunct}
{\mcitedefaultendpunct}{\mcitedefaultseppunct}\relax
\EndOfBibitem
\bibitem[Grimme(2006)]{grim06jcc}
Grimme,~S. Semiempirical GGA-type density functional constructed with a
  long-range dispersion correction. \emph{J.~Comput.~Chem.~} \textbf{2006},
  \emph{27}, 1787--1799\relax
\mciteBstWouldAddEndPuncttrue
\mciteSetBstMidEndSepPunct{\mcitedefaultmidpunct}
{\mcitedefaultendpunct}{\mcitedefaultseppunct}\relax
\EndOfBibitem
\bibitem[Hedin(1965)]{hedi65pr}
Hedin,~L. New method for calculating the one-particle Green's function with
  application to the electron gas problem. \emph{Phys.~Rev.~} \textbf{1965},
  \emph{139}, A796--A823\relax
\mciteBstWouldAddEndPuncttrue
\mciteSetBstMidEndSepPunct{\mcitedefaultmidpunct}
{\mcitedefaultendpunct}{\mcitedefaultseppunct}\relax
\EndOfBibitem
\bibitem[Hybertsen and Louie(1985)Hybertsen, and Louie]{hybe-loui85prl}
Hybertsen,~M.~S.; Louie,~S.~G. First-Principles Theory of Quasiparticles:
  Calculation of Band Gaps in Semiconductor and Insulators.
  \emph{Phys.~Rev.~Lett.~} \textbf{1985}, \emph{55}, 1418--1421\relax
\mciteBstWouldAddEndPuncttrue
\mciteSetBstMidEndSepPunct{\mcitedefaultmidpunct}
{\mcitedefaultendpunct}{\mcitedefaultseppunct}\relax
\EndOfBibitem
\bibitem[Hanke and Sham(1980)Hanke, and Sham]{hank-sham80prb}
Hanke,~W.; Sham,~L.~J. Many-particle effects in the optical spectrum of a
  semiconductor. \emph{Phys.~Rev.~B} \textbf{1980}, \emph{21}, 4656--4673\relax
\mciteBstWouldAddEndPuncttrue
\mciteSetBstMidEndSepPunct{\mcitedefaultmidpunct}
{\mcitedefaultendpunct}{\mcitedefaultseppunct}\relax
\EndOfBibitem
\bibitem[Strinati(1988)]{stri88rnc}
Strinati,~G. Application of the Green Functions Method to the Study of the
  Optical Properties of Semiconductors. \emph{Riv.~Nuovo~Cimento~}
  \textbf{1988}, \emph{11}, 1--86\relax
\mciteBstWouldAddEndPuncttrue
\mciteSetBstMidEndSepPunct{\mcitedefaultmidpunct}
{\mcitedefaultendpunct}{\mcitedefaultseppunct}\relax
\EndOfBibitem
\bibitem[Gulans \latin{et~al.}(2014)Gulans, Kontur, Meisenbichler, Nabok,
  Pavone, Rigamonti, Sagmeister, Werner, and Draxl]{gula+14jpcm}
Gulans,~A.; Kontur,~S.; Meisenbichler,~C.; Nabok,~D.; Pavone,~P.;
  Rigamonti,~S.; Sagmeister,~S.; Werner,~U.; Draxl,~C. exciting: a
  full-potential all-electron package implementing density-functional theory
  and many-body perturbation theory. \emph{J.~Phys.~Condens.~Matter.~}
  \textbf{2014}, \emph{26}, 363202\relax
\mciteBstWouldAddEndPuncttrue
\mciteSetBstMidEndSepPunct{\mcitedefaultmidpunct}
{\mcitedefaultendpunct}{\mcitedefaultseppunct}\relax
\EndOfBibitem
\bibitem[Nabok \latin{et~al.}(2016)Nabok, Gulans, and Draxl]{nabo+16prb}
Nabok,~D.; Gulans,~A.; Draxl,~C. Accurate all-electron $G_0W_0 $
  quasiparticle energies employing the full-potential augmented planewave
  method. \emph{Phys.~Rev.~B} \textbf{2016}, \emph{94}, 035118\relax
\mciteBstWouldAddEndPuncttrue
\mciteSetBstMidEndSepPunct{\mcitedefaultmidpunct}
{\mcitedefaultendpunct}{\mcitedefaultseppunct}\relax
\EndOfBibitem
\bibitem[Sagmeister and Ambrosch-Draxl(2009)Sagmeister, and
  Ambrosch-Draxl]{sagm-ambr09pccp}
Sagmeister,~S.; Ambrosch-Draxl,~C. Time-Dependent Density Functional Theory
  versus Bethe--Salpeter equation: An All-Electron Study.
  \emph{Phys.~Chem.~Chem.~Phys.~} \textbf{2009}, \emph{11}, 4451--4457\relax
\mciteBstWouldAddEndPuncttrue
\mciteSetBstMidEndSepPunct{\mcitedefaultmidpunct}
{\mcitedefaultendpunct}{\mcitedefaultseppunct}\relax
\EndOfBibitem
\bibitem[Gillet \latin{et~al.}(2013)Gillet, Giantomassi, and
  Gonze]{gille+13prb}
Gillet,~Y.; Giantomassi,~M.; Gonze,~X. First-principles study of excitonic
  effects in Raman intensities. \emph{Phys.~Rev.~B} \textbf{2013}, \emph{88},
  094305\relax
\mciteBstWouldAddEndPuncttrue
\mciteSetBstMidEndSepPunct{\mcitedefaultmidpunct}
{\mcitedefaultendpunct}{\mcitedefaultseppunct}\relax
\EndOfBibitem
\bibitem[Kammerlander \latin{et~al.}(2012)Kammerlander, Botti, Marques, Marini,
  and Attaccalite]{kamm+12prb}
Kammerlander,~D.; Botti,~S.; Marques,~M.~A.; Marini,~A.; Attaccalite,~C.
  Speeding up the solution of the Bethe-Salpeter equation by a double-grid
  method and Wannier interpolation. \emph{Phys.~Rev.~B} \textbf{2012},
  \emph{86}, 125203\relax
\mciteBstWouldAddEndPuncttrue
\mciteSetBstMidEndSepPunct{\mcitedefaultmidpunct}
{\mcitedefaultendpunct}{\mcitedefaultseppunct}\relax
\EndOfBibitem
\bibitem[Momma and Izumi(2011)Momma, and Izumi]{momm-izum11jacr}
Momma,~K.; Izumi,~F. {{\it VESTA3} for three-dimensional visualization of
  crystal, volumetric and morphology data}. \emph{J.~Appl.~Cryst.~}
  \textbf{2011}, \emph{44}, 1272--1276\relax
\mciteBstWouldAddEndPuncttrue
\mciteSetBstMidEndSepPunct{\mcitedefaultmidpunct}
{\mcitedefaultendpunct}{\mcitedefaultseppunct}\relax
\EndOfBibitem
\end{mcitethebibliography}
%%%%%%%%%%%%%%%%%%%%%%%%%%%%%%%%%%%
\providecommand{\latin}[1]{#1}
\providecommand*\mcitethebibliography{\thebibliography}
\csname @ifundefined\endcsname{endmcitethebibliography}
  {\let\endmcitethebibliography\endthebibliography}{}

%%%%%%%%%%%%%%%%%%%%%%%%%%%%%%%%%%%%%%%%%%
\includepdf[pages={1,2,3,4,5,6,7,8,9}]{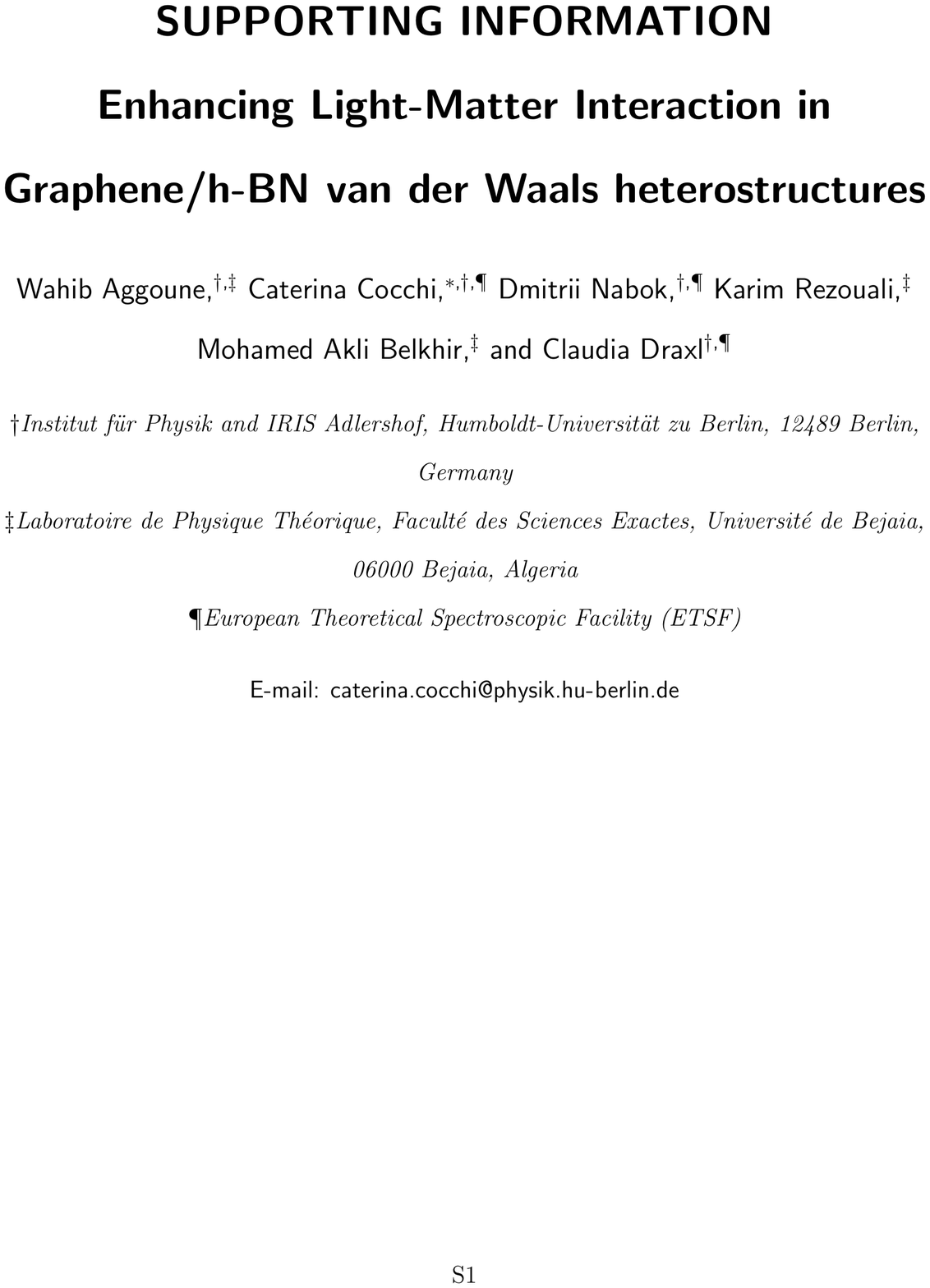}
\end{document}